\documentclass[numbers]{article}
\usepackage{arxiv}

\usepackage[utf8]{inputenc} 
\usepackage[T1]{fontenc}    
\usepackage{hyperref}       
\usepackage{url}            
\usepackage{booktabs}       
\usepackage{amsfonts}       
\usepackage{nicefrac}       
\usepackage{microtype}      
\usepackage{cleveref}       
\usepackage{lipsum}         
\usepackage{graphicx}
\usepackage[font=footnotesize]{caption}
\usepackage[numbers,sort&compress]{natbib}

\usepackage{doi}
\usepackage[separate-uncertainty=true]{siunitx}    

\usepackage[font=scriptsize, labelfont=bf]{caption}

\title{Invasive and Non-Invasive Neural Decoding of Motor Performance in Parkinson's Disease for Personalized Deep Brain Stimulation}

\usepackage{authblk}

\newbox{\orcid}\sbox{\orcid}{\includegraphics[scale=0.06]{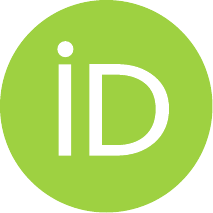}} 
\author[1,2,3]{%
	\href{https://orcid.org/0009-0003-1477-4912}{\usebox{\orcid}\hspace{1mm}Matthias Dold\thanks{\texttt{matthias.dold@donders.ru.nl}}}%
}
\author[3, 4, 5]{%
	\href{https://orcid.org/0000-0002-1703-6283}{\usebox{\orcid}\hspace{1mm}Volker A. Coenen}%
}
\author[3]{%
	\href{https://orcid.org/0000-0002-2836-0046}{\usebox{\orcid}\hspace{1mm}Bastian Sajonz}%
}
\author[3]{%
	\href{https://orcid.org/0000-0003-1691-546X}{\usebox{\orcid}\hspace{1mm}Peter Reinacher}%
}
\author[3]{%
	\href{https://orcid.org/0000-0002-2629-4110}{\usebox{\orcid}\hspace{1mm}Thomas Prokop}%
}
\author[3, 6]{%
\href{https://orcid.org/0000-0003-2742-1940}{\usebox{\orcid}\hspace{1mm}Marco Reisert}%
}
\author[2]{%
	\href{https://orcid.org/0009-0003-0594-7893}{\usebox{\orcid}\hspace{1mm}Sophia Gimple}%
}
\author[7, 8]{%
	\href{https://orcid.org/0000-0002-3589-5604}{\usebox{\orcid}\hspace{1mm}Yasin Temel}%
}
\author[2]{%
	\href{https://orcid.org/0000-0003-0030-7224}{\usebox{\orcid}\hspace{1mm}Marcus L.F. Janssen}%
}
\author[1]{%
	\href{https://orcid.org/0000-0001-6729-0290}{\usebox{\orcid}\hspace{1mm}Michael Tangermann}%
}
\author[3, 5, 1]{%
	\href{https://orcid.org/0000-0002-2032-8981}{\usebox{\orcid}\hspace{1mm}Joana Pereira\setcounter{footnote}{0}\thanks{\texttt{joana.pereira@uniklinik-freiburg.de}}}%
}

\affil[1]{Data-Driven Neurotechnology Lab, Donders Institute for Brain, Cognition and Behaviour, Radboud University, Nijmegen, The Netherlands}
\affil[2]{Department of Clinical Neurophysiology, Maastricht University Medical Center, The Netherlands;  Mental Health and Neuroscience Research Institute, Maastricht University, The Netherlands}
\affil[3]{Department of Stereotactic and Functional Neurosurgery, University of Freiburg - Medical Center, Breisacher Straße 64, 79106, Freiburg, Germany}
\affil[4]{Medical Faculty of the University of Freiburg, Breisacher Str. 153, 79110 Freiburg im Breisgau, Germany}
\affil[5]{Brain-Links Brain-Tools, Intelligent Machine-Brain Interfacing Technology (IMBIT), University of Freiburg, Georges-Köhler-Allee 201, 79110, Freiburg, Germany
}
\affil[6]{Department of Diagnostic and Interventional Radiology, University of Freiburg - Medical Center, Killianstrasse 5a, 79106, Freiburg, Germany}
\affil[7]{Department of Neurosurgery, Maastricht University Medical Center, The Netherlands; Mental Health and Neuroscience Research Institute, Faculty Health, Medicine and Life Sciences, Maastricht University, The Netherlands}
\affil[8]{Istanbul Atlas University, Faculty of Medicine, Istanbul, Turkiye}

\renewcommand{\shorttitle}{CopyDraw scrutinized}

\hypersetup{
pdftitle={CopyDraw scrutinized},
pdfsubject={q-bio.NC, q-bio.QM},
pdfauthor={Matthias Dold, Volker A. Coenen, Bastian Sajonz, Peter Reinacher, Thomas Prokop, Marco Reisert, Sophia Gimple, Yasin Temel, Marcus L.F. Janssen, Jordy Thielen, Michael Tangermann, Joana Pereira},
pdfkeywords={DBS, EEG, ECoG, neural decoding, Parkinson's Disease},
}

\newcommand{\Castano}{Casta{\~n}o-Candamil }
\newcommand{\handmotor}{hand-motor }

\newcommand{\icc}{ICC }
\newcommand{\neuralscore}{neural correlation }
\newcommand{\std}{SD }

\begin{document}
\maketitle

\begin{abstract}


Decoding motor performance from brain signals offers promising avenues for adaptive deep brain stimulation (aDBS) to improve treatment for Parkinson's disease (PD). In a two-center cohort of 19 patients with PD executing a drawing task under time pressure, we decoded \handmotor performance from electroencephalography (n=15) and, critically for clinical translation, electrocorticography (n=4). A total of 35 sessions were recorded. Within each session, patients performed the task multiple times under DBS ON and DBS OFF. Instead of relying on a pre-defined single frequency band, we derived patient-specific biomarkers using a filterbank-based machine learning approach. DBS modulated the kinematics significantly in 23 sessions. Neural decoding of kinematics was possible in 28 sessions (mean Pearson correlation coefficient $\text{r}=0.37$). Our results further demonstrate the modulation of the speed-accuracy trade-offs, with increased drawing speed but reduced accuracy under DBS as the most prevalent variant. A joint evaluation of behavioral and neural decoding outcomes revealed six prototypical scenarios, for which we discuss future personalized aDBS strategies.

\end{abstract}


\section{Introduction}
Parkinson’s disease (PD) is a neurodegenerative disorder~\cite{Bloem2021} characterized by bradykinesia, accompanied by rigidity, tremor, and/or impaired balance~\cite{Postuma2015}, alongside a wide range of non-motor symptoms. Given the heterogeneity of the clinical phenotype and disease progression ~\cite{Horsager2024}, therapeutic management remains challenging. Currently, treatment is offered by dopamine replacement therapies~\cite{Fox2018} and/or deep brain stimulation (DBS)~\cite{deuschl:2006, Hartmann2019, Foote2025}. For DBS, the subthalamic nucleus (STN) or globus pallidus interna (GPi)~\cite{Lachenmayer2021, Au2020, Odekerken2013} are considered the most effective subcortical targets. Most patients undergoing DBS have their systems configured to deliver electrical pulses with fixed parameters, which are applied continuously. Since DBS parameters are typically adjusted by clinical experts during infrequent patient visits (every few weeks or months)~\cite{volkmann:2006, Roediger2023}, short-term fluctuations in stimulation needs are not captured. Given the inherent variability of symptoms, it is reasonable to assume that a more personalized, dynamically adjustable DBS mode would provide optimized therapy. Based on this rationale, adaptive DBS (aDBS) systems have been proposed~\cite{Little2013, Neumann2023}, which use patient-specific biomarkers to regulate stimulation in real time, thereby improving symptom control while reducing side effects~\cite{Guidetti2025}. Various aDBS systems have been implemented in research settings for PD patients~\cite{DeNeeling2026}. These systems modulate DBS amplitude, frequency, or pulse width according to a control strategy, typically based on biomarkers derived from local field potentials (LFPs) recorded at the stimulation site~\cite{Arlotti2021, Velisar2019, Pinafuentes2020}. 
Commercial-grade systems for aDBS have recently become available by Medtronic (Percept, Medtronic, Minneapolis, MN, USA) with a (dual-)threshold approach~\cite{Busch2025} and by Newronika (AlphaDBS, Newronika SpA, Milan, Italy) with a proportional aDBS strategy~\cite{Arlotti2021}, establishing aDBS as a new treatment option~\cite{BronteStewart2025}. 

The above-mentioned approaches use a neurophysiological signal (neural marker) that correlates with the severity of symptoms. The most prominent marker is the beta band power of LFP signals from the STN~\cite{Wijk2023}, which was shown to correlate with motor symptom severity~\cite{Kuehn2008} at the group level. However, a recent multi-center study~\cite{Gerster2025} demonstrated that at the individual level this correlation is unreliable or might even change direction compared to the group level. This underpins the need for a subject-specific evaluation of aDBS strategies.

Clinically, PD symptoms are often assessed with the MDS-UPDRS~\cite{Goetz2008}, but even the motor subscale (part III) requires several minutes of expert evaluation and cannot capture rapid symptom fluctuations. Its ordinal scoring and subjectivity further limit precision~\cite{Hendricks2021}. To address these issues, recent studies investigate behavioral metrics such as direct \handmotor performance, including grip-force~\cite{Herz2023,peterson:2023} and finger tapping~\cite{Lee2025, Gulberti2024, Muehlberg2023}, or reaching tasks~\cite{he:2023, Cavallo2025}. 

A more complex and ecologically relevant movement, similar to micrographia~\cite{Asci2025}, is studied by the CopyDraw task~\cite{Castano2019}. Neural markers have been derived for the CopyDraw task from electroencephalography (EEG) signals using a machine learning (ML) decoding approach~\cite{Castano2020} in a small cohort of PD patients. Using the same task, we have recently reported on patient-specific electrocorticography (ECoG) neural markers in a single patient, which were used for a first proof-of-concept aDBS experiment~\cite{dold2025}. 

These previous studies demonstrated the feasibility of using the CopyDraw task to derive neural markers for aDBS. Here, we extend this work to a larger data set of 19 participants and 35 sessions, to: 
(1) quantify the success rate of the approach with respect to the \handmotor behavior modulation by DBS, and the neural decoding performance on this larger population, 
(2) demonstrate the transferability of the approach to epidural ECoG (n=4, 8 sessions total), as systems for chronic use would need to be fully implanted, 
(3) extend the analysis on \handmotor behavior and neural decoding to a new \handmotor performance measure that describes the trade-off between speed and accuracy made by patients, and how this trade-off is modulated by DBS, and 
(4) derive several scenarios and implications/recommendations for suitable aDBS control strategies based on the joint behavioral and neural decoding outcomes.

\section{Results}

To investigate the electrophysiological correlates of DBS and its effect on \handmotor performance, we analyzed 35 recording sessions from 19 PD patients. During these sessions, patients performed a kinematic tracking paradigm (the CopyDraw task~\cite{Castano2019}) while concurrent neural activity was recorded using either EEG (n=15) or ECoG (n=4). We first characterize the behavioral modulation of \handmotor performance induced by DBS. Building on this, we subsequently identify session-specific neural markers within the EEG and ECoG signals that are predictive of these \handmotor outcomes.

\subsection{DBS modulates behavior during the CopyDraw task}

To determine behavioral differences induced by DBS, we evaluated the prediction of the DBS condition (ON vs. OFF) from a set of behavioral features of the CopyDraw task, including speed, acceleration and jerk. The DBS condition was classifiable from these behavioral features in 23 out of 35 sessions (Figure~\ref{fig:behav_overview} A), manifesting in a significant area under the receiver operating characteristic curve (CopyDraw ROC AUC). The CopyDraw ROC AUC was calculated as the mean across a chronological cross-validation. At the group level, the mean (standard deviation) CopyDraw ROC AUC was 0.68\,($\pm$\,0.13 \std). The mean chance level, based on permutation tests, was $\text{ROC AUC}_{chance}=0.61\,(\pm$\,0.03 \std). When only considering sessions with a significant CopyDraw ROC AUC, the mean CopyDraw ROC AUC was 0.76\,($\pm$\,0.09 \std).

We analyzed the Shapley additive explanation (SHAP)~\cite{Lundberg2017} feature importance values to understand how the \handmotor features were modulated by DBS.
The analysis revealed an pattern for sessions with a higher CopyDraw ROC AUC (Figure ~\ref{fig:behav_overview} B): DBS increased the speed features while acceleration-related features were decreased in magnitude and in the horizontal (x-)direction of the screen. Acceleration in the y-direction was mostly reduced by DBS, but was also increased in a few cases (see S2$_2$ and S6$_c$ in Figure ~\ref{fig:behav_overview} B). No clear patterns pointing towards the relevance of jerk features for the  behavioral decoding were found. To validate the interpretation of the feature importance values, Figures~\ref{fig:behav_overview} C-E show z-scored feature values for selected example sessions. All examples show significant differences (Mann-Whitney U test) between DBS ON and OFF trials. 
No significant differences in CopyDraw ROC AUC were observed between chronic vs.~acute and EEG vs. ECoG sessions (see Figure~\ref{fig:apx_feature_set_session_type_signal_modality}). 

We introduced the \textit{task performance} as a metric for quantifying how well a participant performed the instructed task of copying a target template as fast and as accurately as possible. The task performance is therefore defined as speed, expressed as the fraction of the template that was copied within the given time window, divided by accuracy, expressed by the average distance of the drawn trace and the template.
An example in Figure~\ref{fig:performance_vs_copydraw_score} A illustrates how the task performance is derived by matching points from the target template (blue dots) with points from the drawn trace (orange dots). Gray dots indicate the part of the template that could not be matched by dynamic time warping (DTW)~\cite{Giorgino2009}, as the drawn trace was too short. This effect was typically observed when patients ran into the time-out criterion during the drawing.
By matching the drawn trace with the template, we observed the following: For 17 out of 35 sessions, DBS significantly modulated task performance (Mann-Whitney U test, $P<0.05$, two-sided), see Figure~\ref{fig:performance_vs_copydraw_score} D. We found 7 sessions with a significant increase in task performance and 10 sessions with a significant decrease due to DBS. Looking at the means across all sessions with task performance increases and decreases separately for the two DBS conditions (Figure~\ref{fig:performance_vs_copydraw_score} B), revealed the following structure: Sessions with a decreased task performance were faster on average, to the detriment of accuracy, while sessions with an increased task performance were both faster and more accurate on average.

To understand how the task performance related to the CopyDraw score, we compared the effect size of the task performance (DBS ON vs.~OFF, x-axis) with the CopyDraw ROC AUC (y-axis) in Figure~\ref{fig:performance_vs_copydraw_score} C. Sessions with significant task performance modulation through DBS are indicated by markers with a bold outline. For the CopyDraw ROC AUC, the group average chance level was at 0.61, as indicated with a gray background. Two separate OLS models were fitted, one for sessions that showed a positive effect of DBS ($\text{r}=0.56, P=0.012, \text{R}^2=0.31$), and one for a negative effect ($\text{r}=-0.49, P=0.057, \text{R}^2=0.24$). 

Next, we analyzed how informative the task performance was with respect to the DBS condition. An LDA with the task performance as its single input feature was used to predict the DBS condition (task performance ROC AUC), and was compared to the CopyDraw ROC AUC. The scatter plot in Figure~\ref{fig:performance_vs_copydraw_score} C shows that there are four sessions ($\text{S5}_{3}$, $\text{S7}_{c}$, $\text{S12}_{4}$, $\text{S16}_{4}$) for which the DBS condition is easier to decode from the task performance than from the CopyDraw features of speed, acceleration, and jerk.

\subsection{Decoding from neural signals}

\subsubsection{CopyDraw scores can be predicted from neural signals}
After quantifying the behavioral effects of DBS, the next step is to understand whether the \handmotor behavior can be predicted from neural data. If we can predict the behavior from neural signatures, we can consider them as relevant neural markers.
We report a significant prediction of the CopyDraw score from neural data in 28 (out of 35) sessions (Figure~\ref{fig:neural_overview} A). At the group level, the mean neural correlation is $\text{r}=0.37\,(\pm\,0.23\,\text{\std})$ with mean chance level at $\text{r}_{chance}=0.20\,(\pm\,0.06\,\text{\std})$. Note that a significant decoding of motor behavior was achieved in six sessions ($\text{S5}_3$, $\text{S6}_1$, $\text{S7}_c$, $\text{S11}_c$, $\text{S12}_2$, $\text{S12}_4$) for which the \handmotor behavior was not significantly modulated by DBS (CopyDraw ROC AUC not significant). The opposite scenario (significant CopyDraw ROC AUC and no significant neural decoding) is found for one session only ($\text{S8}_c$). 

To investigate which features were relevant for the neural decoding, we analyzed our pipelines, which used a filter bank version of the source-power comodulation (SPoC)~\cite{Dahne2014} spatial filtering.  For this purpose, we counted the number of spatial features per frequency band as selected by the minimum redundancy maximum relevance (MRMR)~\cite{Peng2005} algorithm. These counts reveal a roughly even distribution across the frequency bands (Figure~\ref{fig:neural_overview} B), with $\text{N}=58$ for ${\vartheta}$ (4--8)\,Hz, $\text{N}=54$ for $\alpha$ (8--12)\,Hz, $\text{N}=53$ for $\beta$ (12--30)\,Hz, $\text{N}=60$ for $\gamma$ (30--45)\,Hz, and $\text{N}=55$ for $\gamma_{high}$ (55--90)\,Hz. Individual sessions show high variability, with, e.g., strong focus on beta for $\text{S7}_3$ and $\text{S6}_2$, while $\text{S14}_4$ or $\text{S1}_2$ do not select any feature from the beta frequency range. Selected stereotypical features for sessions with successful neural decoding of the CopyDraw score are provided in Figure~\ref{fig:marker_overview}. All features for the pipelines of the selected sessions can be found in the appendix (Figures~\ref{fig:apx_eeg_marker_S12_2},~\ref{fig:apx_eeg_marker_S10_c},~\ref{fig:apx_eeg_marker_S1_c},~\ref{fig:apx_ecog_marker}).

\subsubsection{DBS condition is predictable from neural signals}
A biomarker, i.e., in our case, the output of our neural regression pipelines, needs to be modulated by DBS to be suitable for informing an aDBS system. We used the features found by the regression pipelines---frequency and spatially filtered band powers---as input to a classifier predicting the DBS condition. If DBS is modulating the features, we should be able to decode the DBS condition from them. The results are presented in Figure~\ref{fig:neural_overview} C. Significant decoding was achieved for 26 (out of 35) sessions, with group-level mean (regression features) ROC AUC = $0.71\,(\pm\,0.15\,\text{\std})$ and mean permutation chance level at $0.65\,(\pm\,0.06\,\text{\std})$. Five sessions ($\text{S1}_3$, $\text{S11}_c$, $\text{S12}_4$, $\text{S13}_c$, $\text{S14}_4$) achieved a significant neural correlation with the CopyDraw score, but the extracted features could not be used to classify the DBS condition. 

\subsubsection{Decoding task performance from neural signals}
When using the neural regression pipeline to predict the task performance instead of the CopyDraw score, the mean neural correlation is significantly (Welch's t-test, $P=7\cdot10^{-6}$) lower, see Figure~\ref{fig:neural_overview} D. However, both decoding performances show a clear linear relation ($\text{r}=0.44, P=0.0087, \text{R}^2=0.19$, see Figure~\ref{fig:neural_overview} E). In all but seven sessions ($\text{S2}_{c}$, $\text{S3}_{c}$, $\text{S4}_{2}$, $\text{S7}_{2}$, $\text{S8}_{c}$, $\text{S11}_{c}$, $\text{S16}_{4}$), neural data correlate more with the CopyDraw score than with the task performance.

\subsection{Prototypical outcome types}
\begin{figure}
    \centering    
    \includegraphics[width=\linewidth]{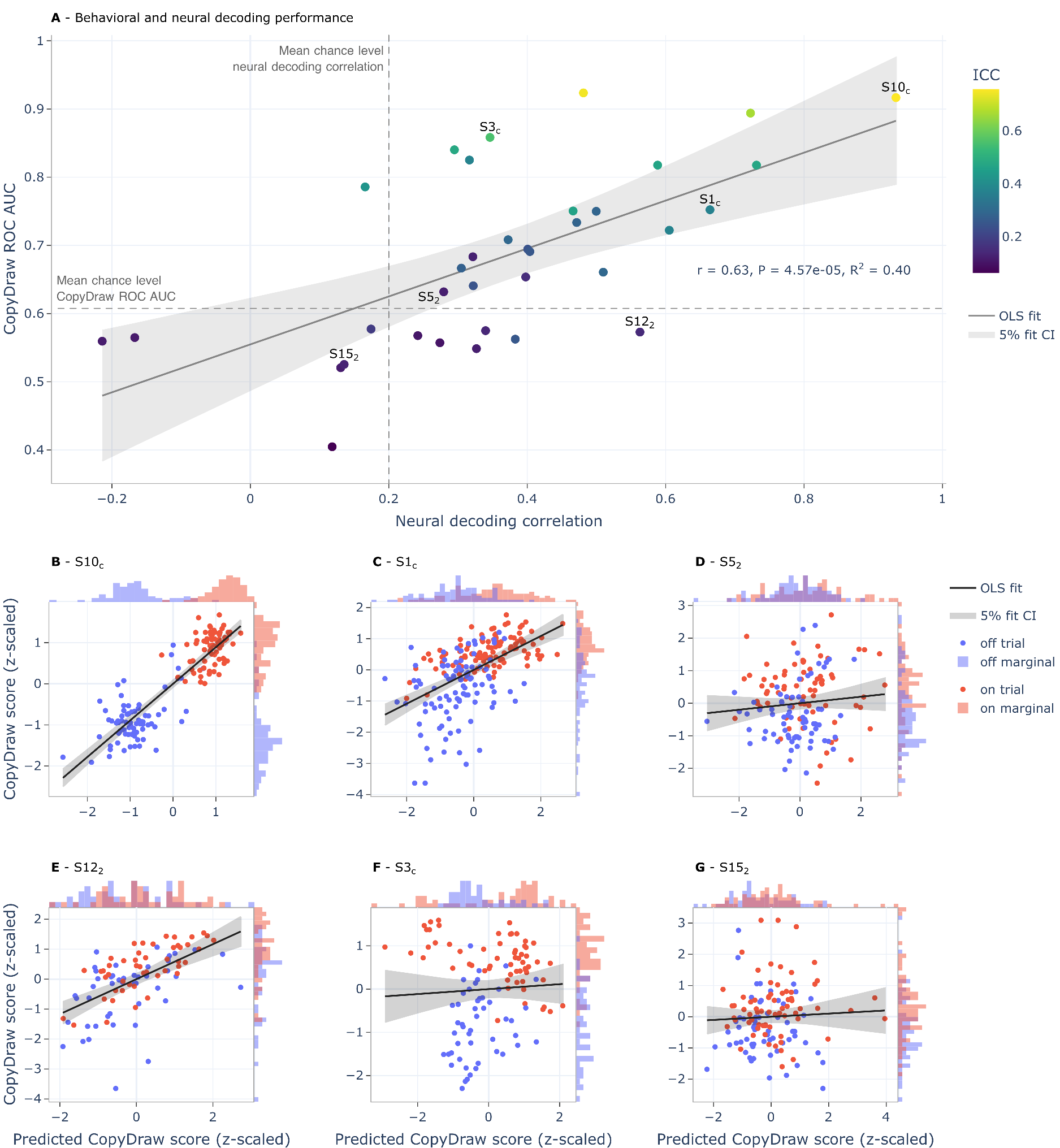}      
    \caption{Correlation between behavioral and neural decoding and examples of different outcome types. \textbf{A} - Correlation scatter between CopyDraw ROC AUC (y-axis) and mean correlation from neural decoding (x-axis). Ordinary least squares (OLS) fit shows a significant correlation with $\text{r}=0.63$ ($P<0.001$) with $\text{R}^2=0.4$ and is drawn as a dark gray line with 5\,\% confidence intervals (CI, gray area). Scatter points are colored according to the intracluster correlation coefficient (ICC). Sessions marked with a text label are presented as stereotypical examples in B-G. \textbf{B}-\textbf{G} visualizes the neural decoding correlations as used in A. Each scatter plot shows the actual CopyDraw scores (y-axis) vs.~the CopyDraw score predictions from neural data (x-axis). Each point refers to a single trial and is colored according to the DBS condition---red for DBS ON and blue for DBS OFF. 
    Like in A, the fitted OLS regression models are drawn as black lines. Marginal distributions are plotted as normalized histograms to the corresponding axes, colored according to the DBS condition. 
    \textbf{B} shows all trials of $\text{S}10_c$---strong DBS effect on the CopyDraw score, which is also captured in the neural predictions. Both marginals show a clear bimodal distribution, indicating a high ICC. \textbf{C} shows all trials of $\text{S}1_c$---an example of a continuous modulation of the CopyDraw score by DBS with good neural decoding performance. \textbf{D} shows all trials of $\text{S}5_2$---an example for a DBS modulation of the CopyDraw score, which still contains a lot of variability beyond the DBS modulation. No significant neural decoding was possible for this session. \textbf{E} shows all trials of $\text{S}12_2$---an example for a session in which no significant modulation of the CopyDraw score was possible, but in which the CopyDraw score could be predicted from neural signals. \textbf{F} shows all trials of $\text{S}3_c$---an example of a session in which a strong DBS effect on the CopyDraw score was observed, but which could not be decoded from neural signals. \textbf{G} shows all trials of $\text{S}15_2$---no effect of DBS on the CopyDraw score was observed, and no neural decoding was possible.}
    \label{fig:outcome_types}
\end{figure}

First, the CopyDraw score was derived from a decoding pipeline using behavioral data with the DBS condition as labels. Next, the CopyDraw score was used as a label in a decoding pipeline using neural signals to identify neural markers. To understand the dependency between the decoding performances of these two decoding pipelines, we compared the neural decoding correlation (Pearson's r for predicting the CopyDraw score, x-axis) and CopyDraw ROC AUC (y-axis) in Figure~\ref{fig:outcome_types} A. Each session is represented as a scatter marker and colored according to the ICC of the CopyDraw score to highlight a potential bimodal distribution. 
On the group level, and as expected, CopyDraw ROC AUC and neural decoding correlation show a linear relation with $\text{r}=0.63, P=4.57\cdot10^{-5}, \text{R}^2=0.40$, see Figure~\ref{fig:outcome_types} A. 

When considering the CopyDraw score and predictions from the neural decoding model, six different outcome types were found. Sessions selected as stereotypical examples (text labels in Figure~\ref{fig:outcome_types} A) are provided in Figure~\ref{fig:outcome_types} B-G with their true (y-axis) and predicted (x-axis) CopyDraw scores per trial: $\text{S10}_c$ (Figure~\ref{fig:outcome_types} B), a session with a strong separation in both the CopyDraw score and the neural prediction with bimodal marginal distributions; $\text{S1}_c$ (Figure~\ref{fig:outcome_types} C), a session with good CopyDraw score separation and neural prediction, but without bimodal marginals; $\text{S5}_2$ (Figure~\ref{fig:outcome_types} D), a session with CopyDraw scores showing separation in DBS ON and OFF but with failed neural prediction;  $\text{S12}_2$ (Figure~\ref{fig:outcome_types} E), a session with a significant neural prediction of the CopyDraw score, but without significant DBS effect on the CopyDraw score; $\text{S3}_c$ (Figure~\ref{fig:outcome_types} F), a session with strong DBS effect on the CopyDraw score (bimodal marginals), but without significant neural prediction; $\text{S15}_2$ (Figure~\ref{fig:outcome_types} G), a session with neither DBS effect, nor significant neural prediction.

\begin{figure}
    \centering   
    \includegraphics[width=\linewidth]{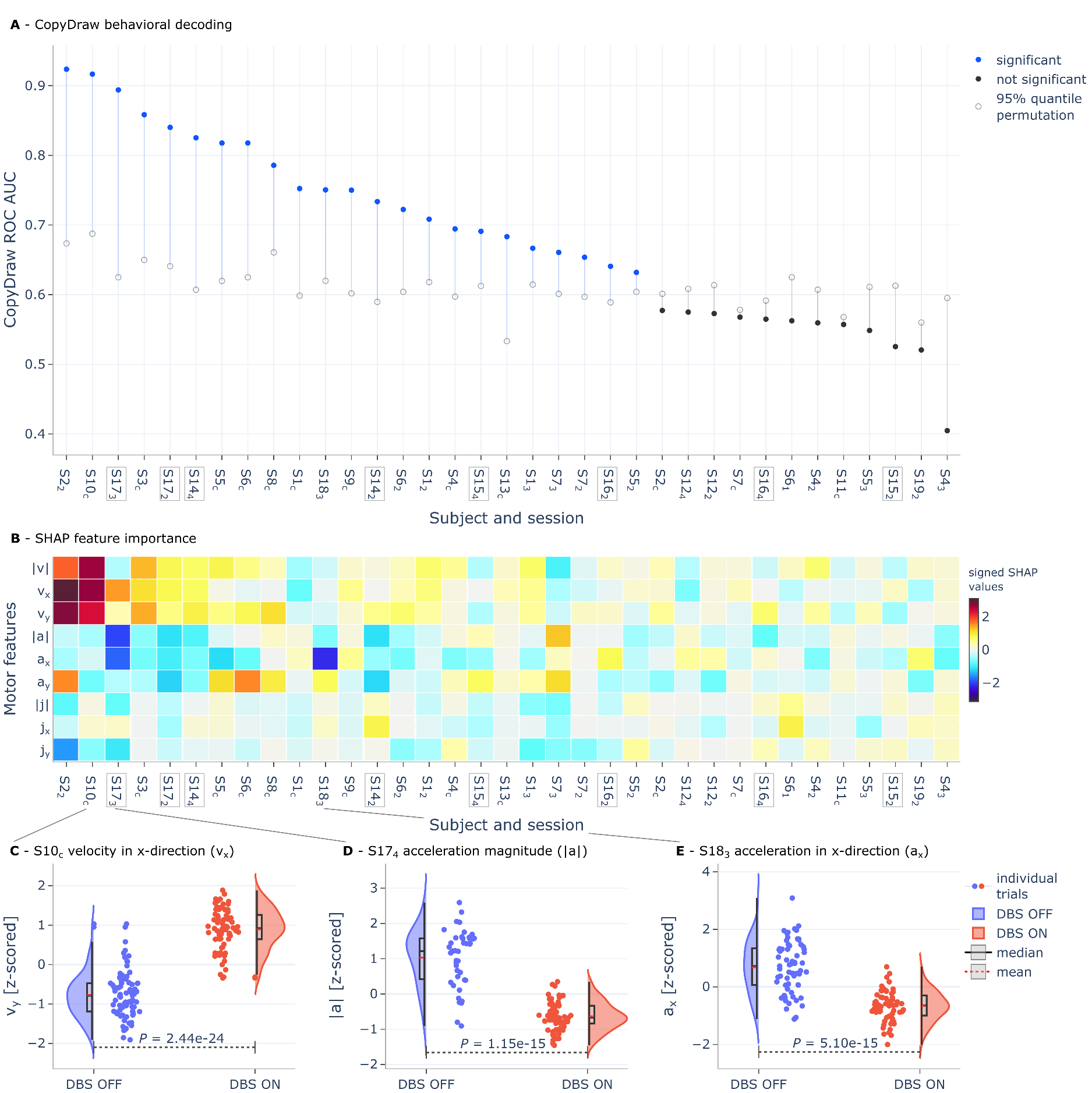}      
    \caption{Decoding DBS condition from behavior - CopyDraw ROC AUC. \textbf{A}: Behavioral decoding performance per subject and session (x-axis, session in subscript, numeric values correspond to days after surgery, subscript ``c'' stands for chronic sessions), sorted according to the CopyDraw ROC AUC (mean across LDA predictions in chronological cross-validation). Markers are colored by significance according to N=1000 bootstrap permutations. 95\,\% chance percentiles are shown with hollow markers. Session codes surrounded by a frame indicate sessions with invasive ECoG data. \textbf{B}: Signed SHAP values (feature importance) for LDA models trained on all data of the corresponding sessions, sorted according to A. Positive SHAP values indicate larger features under DBS ON.
    \mbox{\textbf{C - E}}: Visualization of selected features and sessions to help with the interpretation of SHAP values. Features are presented as they are used within the LDA (clipped to three standard deviations per DBS condition and z-scored). All differences are significant (Mann-Whitney U test, two-sided). Abbreviations: ROC AUC, area under the receiver operating characteristic curve; SHAP Shapley, additive explanation.
    }
    \label{fig:behav_overview}
\end{figure}

\begin{figure}
    \centering 
    \includegraphics[width=\linewidth]{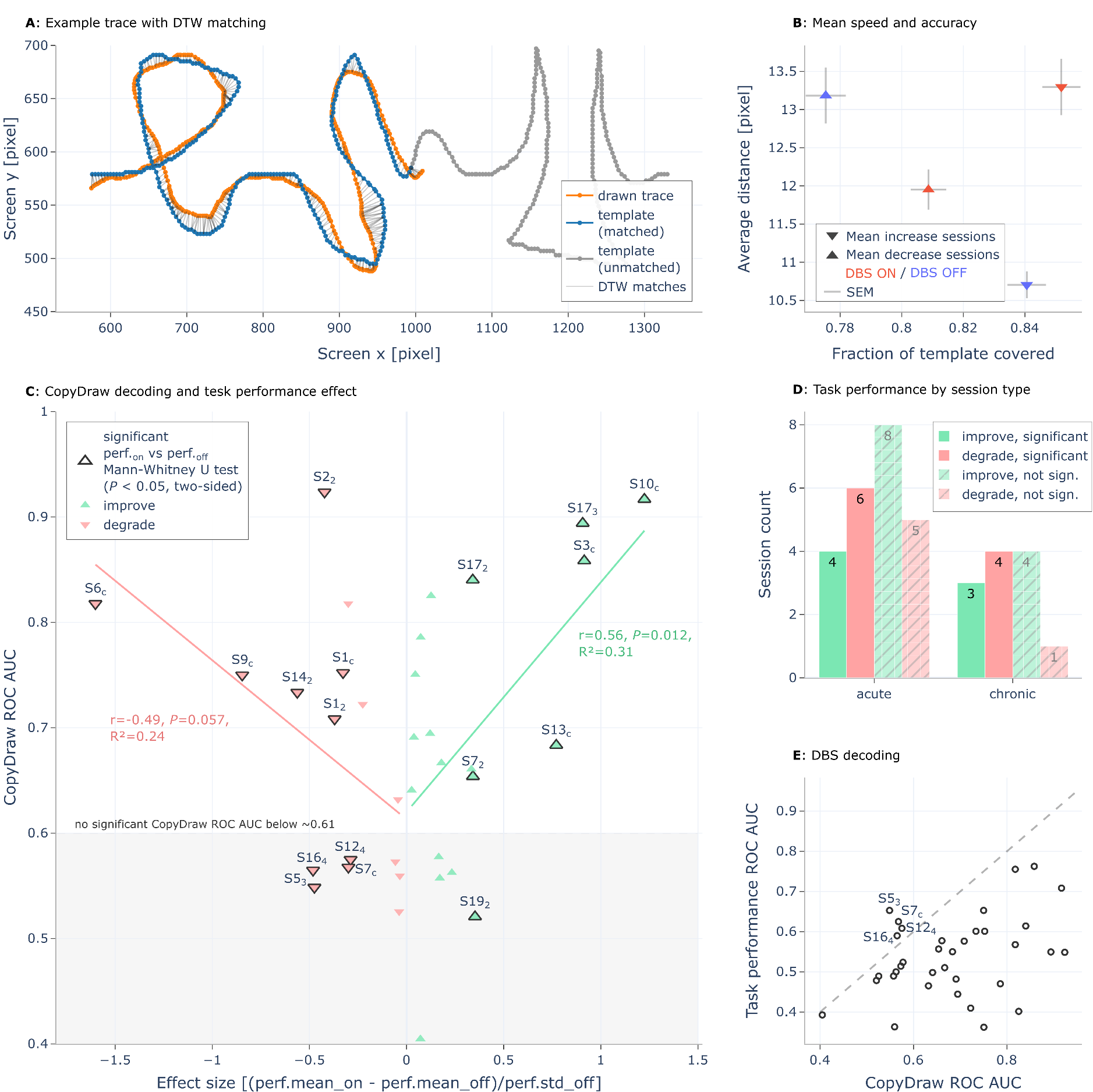}      
    \caption{Task performance evaluation. \textbf{A}: Example template and trace visualizing components relevant for the task performance calculation. The target template (what should be drawn) is shown in blue for the parts that were matched by dynamic time warping (DTW) to the participants' actual drawing (orange), and unmatched parts are shown in gray. Small gray lines visualize how trace samples are matched to template samples. \textbf{B}: Mean fraction of template covered (x-axis, the ratio of blue points to total template points in A), and mean average distance between trace and template. The means are calculated for each DBS condition and across all sessions which increased or decreased in task performance due to DBS separately. Gray lines indicate the $\pm$ one standard error of the mean (SEM) range. \textbf{C}: Scatter plot comparing the DBS effect size on task performance (x-axis) with the CopyDraw ROC AUC (y-axis). Sessions with a significant effect size (Mann-Whitney U test, two sided) are shown with a black outline and text labels. Ordinary least squares (OLS) fit lines are added for all sessions with positive effect size (increase under DBS) and all sessions with negative effect size (decrease under DBS) separately. Only the positive effect size side shows a significant correlation with r$=0.56$ ($P=0.012$) with $\text{R}^2=0.31$ while the negative side is not significant with r$=-0.49$ ($P=0.057$) with $\text{R}^2=0.24$. \textbf{D}: Session counts separated in positive (increase) and negative (decrease) effect size, separately for acute and chronic sessions. Sessions with no significant effect are shown with reduced opacity. \textbf{E}: Scatter plot comparing behavioral decoding of the DBS condition from CopyDraw score (x-axis) and from the same pipeline receiving only the task performance as a feature (y-axis). The dashed line shows x=y. Text labels are provided for sessions in which the task performance feature is more informative than the CopyDraw behavioral features for decoding of the DBS condition. Abbreviations: ROC AUC, area under the receiver operating characteristic curve;}
    \label{fig:performance_vs_copydraw_score}
\end{figure}

\begin{figure}
    \centering   
    \includegraphics[width=\linewidth]{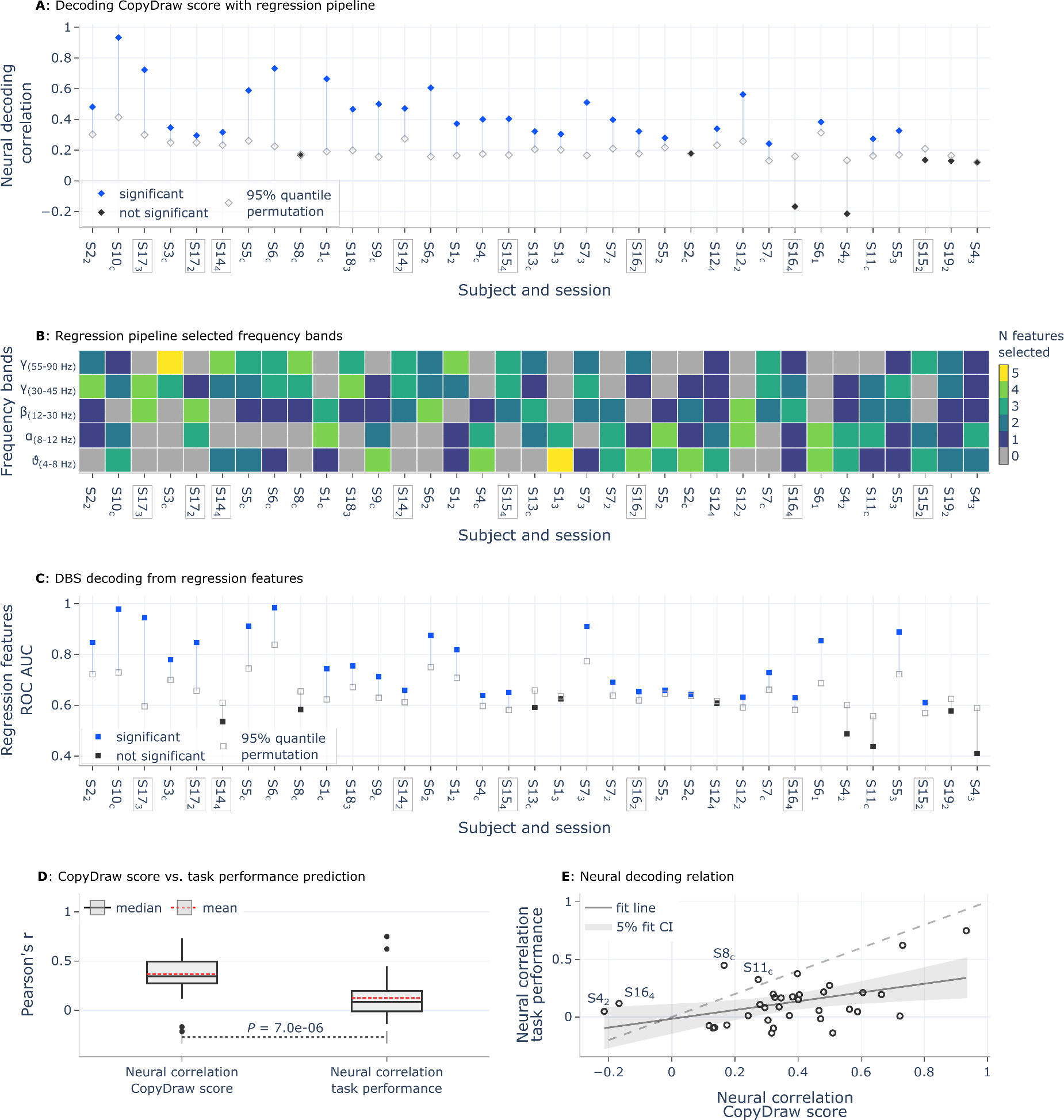}      
    \caption{Neural decoding. \textbf{A}: Neural decoding of the CopyDraw score (diamond markers, y-axis) per subject and session (x-axis, session in subscript, numeric values correspond to days after surgery, subscript ``c'' stands for chronic sessions). Markers are colored by significance according to N=1000 bootstrap permutations. 95\,\% chance percentiles are shown with hollow markers. Sessions are sorted according to CopyDraw ROC AUC (see Figure \ref{fig:behav_overview}). Tick labels with a frame correspond to sessions with invasive ECoG data. \textbf{B}: Counts of selected relevant features per frequency band. Features were selected using the Minimum Redundancy Maximum Relevance (MRMR) feature selection algorithm. Sessions are sorted according to A. \textbf{C}: Performance of using the features (frequency and spatial filters) found in the regression pipeline (used in A) for the classification of the DBS condition (regression feature ROC AUC). Sessions are sorted as in A, with markers colored by significance according to N=1000 bootstrap permutations. 95\,\% chance percentiles are shown with hollow markers. \textbf{D}: Box plots for Pearson's correlation using the regression pipeline to predict the CopyDraw score vs.~using the regression pipeline to predict the task performance. Both pipelines are evaluated using the chrono-CV with the corresponding target variable (CopyDraw score or task performance). Decoding the CopyDraw score achieves significantly higher average correlation ($P=7\cdot10^{-6}$). \textbf{E}: Scatter plot for neural correlation with the CopyDraw score (x-axis) and with the task performance (y-axis). The dashed line shows x=y. Text labels are provided for sessions in which the task performance can be decoded with higher correlation than the CopyDraw score. An ordinary least squares (OLS) fit shows significant correlation with r$=0.44$ ($P=0.0087$), $\text{R}^2=0.19$, see \textit{OLS fit} with 5\,\% confidence intervals (CI). Abbreviations: FBSPoC, filterbank source power comodulation; ROC AUC, area under the receiver operating characteristic curve; CV, cross-validation.}
    \label{fig:neural_overview}
\end{figure}

\begin{figure}
    \centering   
    \includegraphics[width=\linewidth]{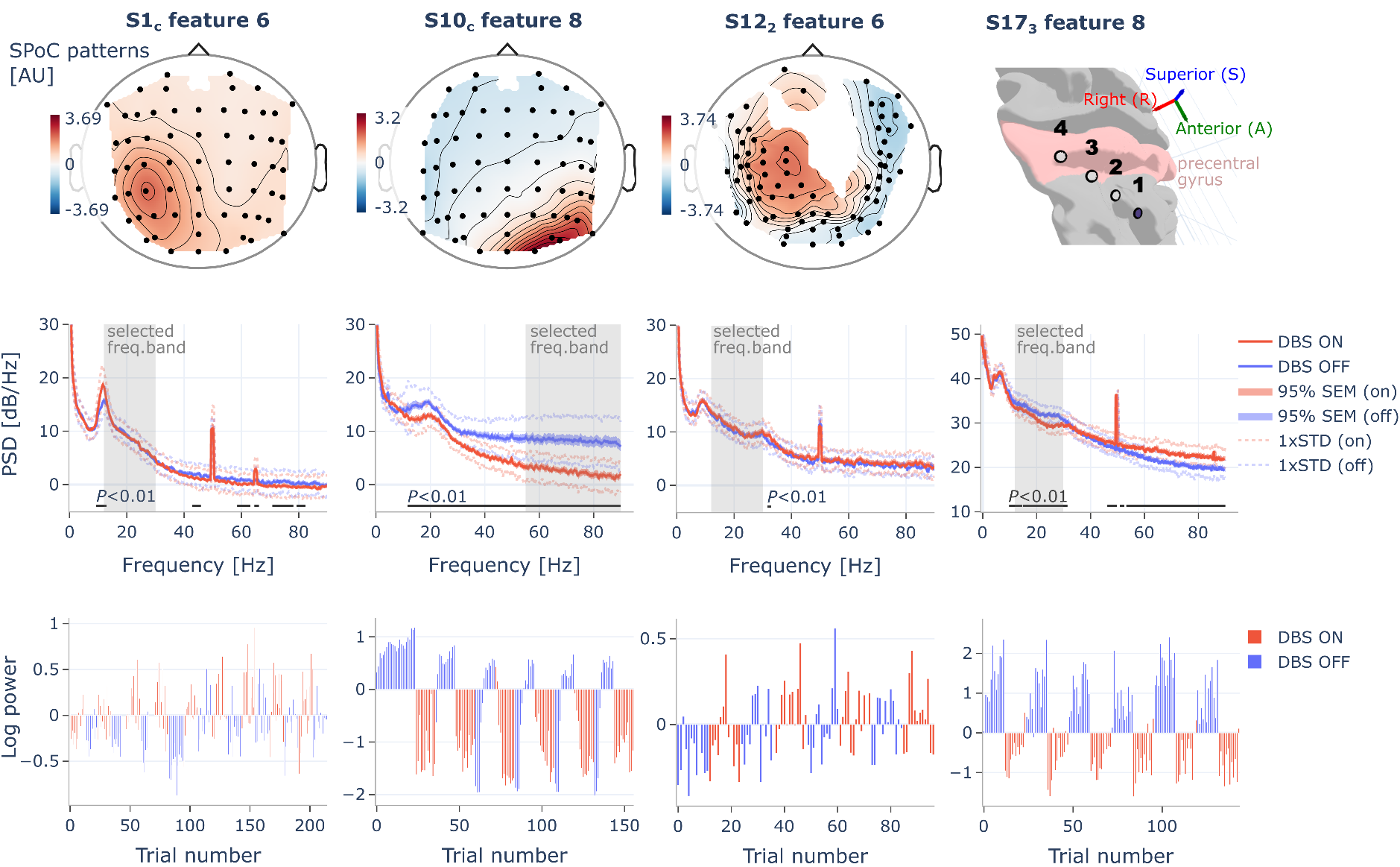}      
    \caption{Selected features showcasing features used for predicting the CopyDraw score. The top row shows spatial patterns (in arbitrary units - AU) for three sessions with EEG and the location of the four contact ECoG strip for $\text{S17}_3$ over the Freesurfer average (fsaverage) brain, with the precentral gyrus highlighted. For the acute session ($\text{S12}_2$), the interpolation of the heatmap excludes the surgical wound area for which no EEG electrodes could be applied. The second row shows broadband power spectral density (PSD) for the spatially filtered signal (filters associated with the patterns shown above). For the ECoG feature, data corresponds to channel 1. Data is shown as mean across trials for DBS ON and DBS OFF, with colored background for the standard error of the mean (SEM) and dashed outlines for +/- one standard deviation (\std). Gray backgrounds highlight the frequency range on which the spatial filters were fitted. Cluster permutation tests were applied with significance at $P<0.01$ shown by black horizontal lines. The third row showcases the scaled band power of the spatially and frequency filtered signals as bars for each trial, colored according to DBS ON and OFF. These values are the input to the final linear regression layer in the decoding pipelines.}
    \label{fig:marker_overview}
\end{figure}

\begin{figure}
    \centering    
    \includegraphics[width=\linewidth]{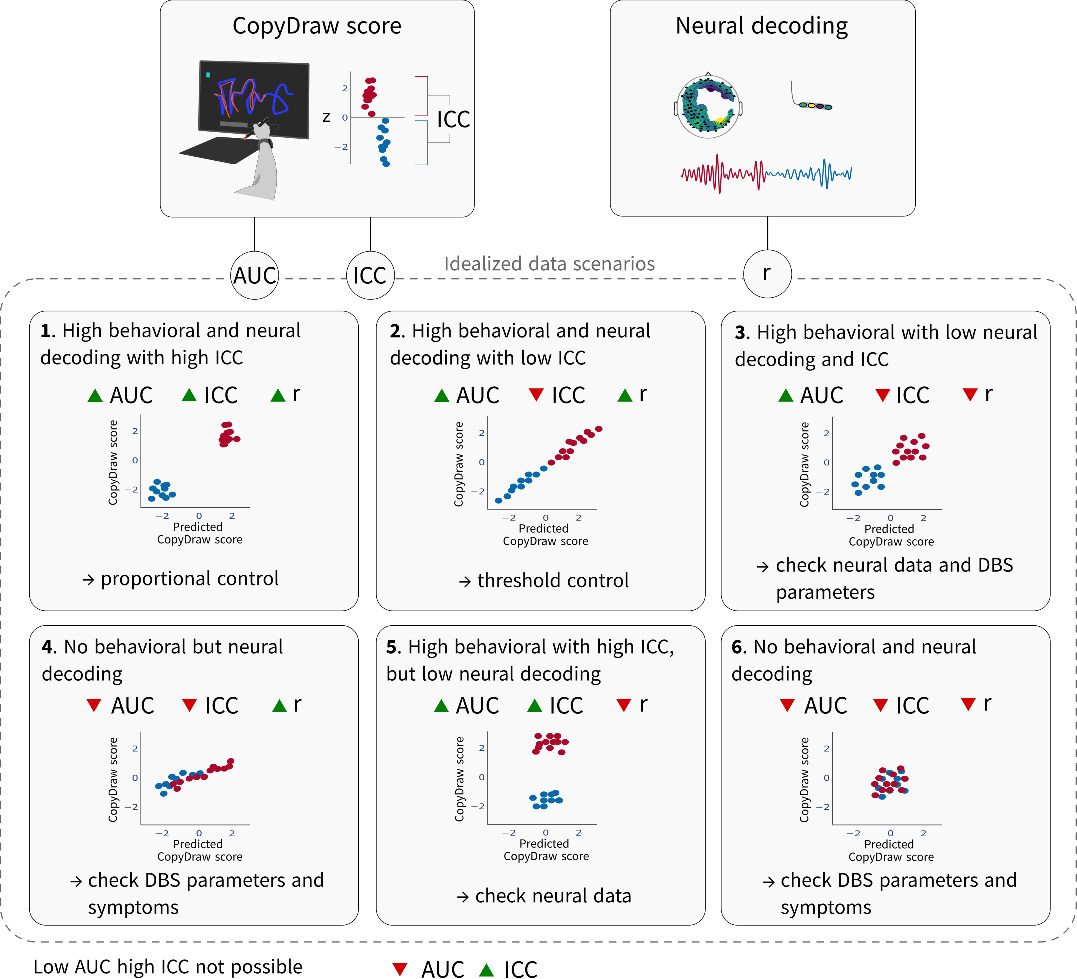}      
    \caption{Possible outcomes of behavioral decoding (CopyDraw ROC \textbf{AUC}, and intracluster correlation coefficient \textbf{ICC}) and neural decoding (Pearson's \textbf{r}) of the CopyDraw score and derived consequences for the usability of markers for aDBS. \textbf{1 - 6} show hypothetical data for idealized scenarios, aligned to the real data scenarios provided in Figure~\ref{fig:outcome_types} B-G.
    \textbf{1}. High AUC, ICC and r make up sessions suitable for proportional control. \textbf{2}. High AUC, low ICC and high r suit a threshold type control. \textbf{3}. High AUC with low ICC and r suggest either a too subtle DBS effect, or suboptimal neural decoding, potentially due to data quality issues. 
    \textbf{4}. Low AUC and ICC with high r, suggest a missing DBS effect, either caused by suboptimal stimulation settings or ceiling effect on \handmotor symptoms (nothing to improve/modulate). \textbf{5}. High AUC and ICC with low r, i.e., failed neural decoding with strong behavioral effect, suggest potential data issues or a suboptimal decoding pipeline. \textbf{6}. Low AUC, ICC and r values suggest checking the DBS parameter or a potential ceiling effect of symptoms first, as a behavioral effect is usually necessary for neural decoding to be successful (see correlation in Figure~\ref{fig:outcome_types}). The derivation of the consequences is detailed in the discussion section~\ref{sec:discussion_scenarios}. A combination of low AUC and high ICC is not possible, as a high ICC implies a clear linear separability.}
    \label{fig:outcome_types_schema}
\end{figure}

\section{Discussion}
\label{discussion}

The present study systematically evaluated the CopyDraw drawing task as a behavioral and neural decoding tool. First, we demonstrated DBS-induced behavioral modulation, as captured by two different scores. These are the CopyDraw score~\cite{Castano2019} and a new task performance metric, which offers a complementary perspective capturing patient-specific trade-offs between drawing speed and accuracy. Our findings on the task performance reveal that DBS did not generally "improve" motor performance in the CopyDraw task, pointing to complex interactions between DBS, motor execution, and higher-order aspects of motor strategies typical of ecological settings. Second, neural decoding of behavioral scores was possible from EEG in both acute and chronic recording sessions. Additionally, our results show transferability of the approach to epidural ECoG recordings (in a small sample), which is particularly important from a translational standpoint for future fully-implantable devices. Finally, the joint analysis of behavioral and neural decoding correlation revealed a structured landscape of outcome types. For each of these outcomes, we provide actionable guidance for the implementation of aDBS control strategies.

\subsection{Main findings}
\subsubsection{\textbf{Decoding Deep Brain Stimulation condition from behavioral signals}}

Observing a significant DBS effect on the CopyDraw score in 23 out of 35 sessions (66\,\%, Figure \ref{fig:behav_overview} A), a canonical question is if this response rate is to be expected. With studies usually reporting on the group-level efficacy of DBS \cite{Starr2025, Giannini2025, he:2023}, it is difficult to find individual or session-specific response rates, especially for the acute phase. 
Partially, this can be explained by the experimental design, which typically collects behavioral or clinical measurements only once under different DBS and/or medication conditions~\cite{Giannini2025}. Significance would then need to be assessed using an a priori defined threshold, e.g., 30\,\% improvement in MDS-UPDRS III~\cite{CastilloTorres2025}, between the different conditions. To not have to impose such thresholds ex-post, we compare the success rate to more longitudinal studies. Recent literature on the twelve month follow-up reports response rates which are broadly in agreement with what we observe in our mixed cohort of acute and chronic measurements. Examples are significant changes in MDS-UPDRS severity levels with $\approx$80\,\% response rate for MDS-UPDRS parts III and IV~\cite{CastilloTorres2025}, 74\,\% response rate for MDS-UPDRS II score improvement~\cite{Morgante2025}.

Our analysis of the SHAP feature importances (Figure \ref{fig:behav_overview} B) reveals that, in general, sessions with a strong DBS effect on the CopyDraw score show higher velocity and lower acceleration values during DBS ON. This suggests a smoother movement execution, which is consistent with observations of increased smoothness in a spiral drawing task linked to improvements in rigidity and bradykinesia~\cite{Radmard2021}. Some sessions (most notably $\text{S2}_{\text{2}}$ and $\text{S6}_{\text{c}}$) show a deviation from this pattern, with a higher acceleration in the y-direction ($\text{a}_{\text{y}}$) under DBS ON. A possible reason for the different impact of DBS on acceleration in the x- and y-directions could be explained by the characteristics of the pseudo-letter templates of CopyDraw. The templates require changes in the y-direction (11 turning points) more frequently than in the x-direction (8 turning points).

\subsubsection{\textbf{Task performance modulation under Deep Brain Stimulation}}
Considering modulation of the task performance, we observed that DBS more often leads to a significant \textit{decrease} in task performance (10 sessions with decrease vs.~7 sessions with increase). This holds for both acute and chronic sessions, see inset in Figure \ref{fig:performance_vs_copydraw_score} A. As the CopyDraw task requires making a trade-off between accuracy and speed, it not only captures \handmotor execution alone, but also higher-level motor planning and aspects of a patient's mental strategy. A possible explanation for the decreased task performance under DBS ON could be a more carefree strategy, which is less focused on precision. This would be in line with changes in impulsivity and risk-taking reported as side effects of DBS~\cite{Herz2024, Scherrer2020, Paliwal2019}. When looking separately at the accuracy and the amount of the template drawn (Figure~\ref{fig:performance_vs_copydraw_score} B and appendix Figure~\ref{fig:apx_task_perf_speed_vs_accuracy}), we observe that sessions with a declining DBS effect mostly (8 out of 10 sessions) show a reduced accuracy under DBS ON. The opposite is observed for sessions with an improving DBS effect (5 out of 7 sessions show a higher accuracy under DBS ON). 
This finding links to a broader discussion of how DBS treatment achieves clinical efficacy. 
E.g., a recent study has characterized PD as a dysfunction of the somato-cognitive action network (SCAN)~\cite{Ren2026}. The authors show that effective DBS modulates SCAN, which is involved in motor planning and behavioral motivation. Such a modulation of motivation would explain cases like $\text{S6}_c$, $\text{S7}_c$, and $\text{S9}_c$, for which only the accuracy decreased under DBS, while the participant copied the whole trace on average in both DBS conditions. That motor-state dependent DBS is modulating brain-wide network activity~\cite{Cavallo2025}, mainly through basal-ganglia and supplementary motor area, is further evidence for the complex nature of the DBS effect.
Considering purely the motor capacity, a decline in task performance could have one of two causes: (1) DBS is having a negative impact on capabilities, e.g., driving the patient into a hyperkinetic state, resulting in difficulties performing the task. Or (2), DBS is modulating capabilities, e.g., allowing the participant to draw precisely at all. This new capability might then result in the participant focusing more on accuracy at the cost of speed. For participants who became more accurate under DBS ON at the cost of reduced speed (see appendix Figure~\ref{fig:apx_task_perf_speed_vs_accuracy}, subjects $\text{S1}_c$, $\text{S12}_4$, $\text{S17}_2$, $\text{S17}_3$), we cannot unravel if this is a change in mental strategy or whether the new found ability to be accurate in the first place (due to DBS) leads to a stronger focus on accuracy. Furthermore, the same decision (being more accurate but slower) can lead to an increase in task performance ($\text{S17}_2$, $\text{S17}_3$) or a decrease ($\text{S1}_c$, $\text{S12}_4$), depending on where the participant is located in the task performance domain (appendix Figure~\ref{fig:apx_task_perf_speed_vs_accuracy} A) and how large the change is. As participants did not receive feedback about their task performance, optimization of the accuracy vs.~speed trade-off is up to an uninformed personal choice. Such a freedom of choice is not idiosyncratic to our task, but  is typical for ecological tasks, as, e.g., within the finger tapping assessment of the MDS-UPDRS, and remains a challenge for assigning a performance metric.\\
Comparing the task performance with the CopyDraw ROC AUC shows a correlation in magnitude (Figure \ref{fig:performance_vs_copydraw_score} C). The correlation coefficients have different signs and while correlation for sessions with an increasing performance effect ($\text{r}=0.56$, $P=0.012$) is significant, it falls short of significance for sessions with a decreasing performance effect ($\text{r}=-0.49$, $P=0.057$). This highlights that a high CopyDraw score cannot be interpreted directly as an accurate or fast replication of the template. The CopyDraw score was designed to capture a linear combination of \handmotor features, which is most clearly separated by DBS~\cite{Castano2019}. Together with the sign convention that, on average, CopyDraw scores are greater than zero under DBS ON, it is plausible that the correlation exists only in magnitude. Additionally, we would expect the DBS effect to be more clearly visible in the CopyDraw score---due to the classification model, which is trained on this effect---compared to the task performance. This is confirmed for all but four sessions, see Figure~\ref{fig:performance_vs_copydraw_score} C.

\subsubsection{\textbf{Decoding CopyDraw scores from neural signals}}
Regarding the identification of neural markers for the CopyDraw score, we observed a good success rate of the proposed pipeline, with significant decoding in 28 out of 35 sessions (80\,\%, Figure~\ref{fig:neural_overview} A). The average neural decoding correlation of 0.37 is on the same level as the results reported on the subset of sessions by \Castano et al.~\cite{Castano2020}, which required a pipeline with optimized narrow-band filters with fine-tuning for each session.  

For 23 of these 28 sessions, the features of the pipelines could also be used for a significant decoding of the DBS condition (Figure~\ref{fig:neural_overview} C), hence the corresponding markers, i.e., the predicted CopyDraw scores, are controllable by DBS. 

To better understand how these pipelines manage to decode the CopyDraw scores, Figure~\ref{fig:marker_overview} illustrates four examples of features used in these pipelines. All four lead to a successful neural decoding, but for partially different underlying reasons.\\
Session $\text{S1}_c$ (Figure~\ref{fig:marker_overview}, first column) shows a left centro-parietal pattern, with a significantly increased spectral power in the alpha band, just below the selected frequency range (13--30\,Hz), for DBS ON. This is an example in which a threshold-based aDBS system~\cite{Little2013}, which turns on stimulation when observing power values above the threshold, would be stuck in the DBS ON state. Assuming a therapeutic benefit under DBS ON, this would align with findings of reversed correlation between band power and symptom severity~\cite{Gerster2025}. Thus, when interpreting the power values correctly for this individual patient, this feature may be suitable for future aDBS evaluation.\\
Session $\text{S10}_c$ (Figure~\ref{fig:marker_overview}, second column) shows a right-occipital spatial pattern. This component significantly attenuates in beta- and gamma band power under DBS ON. It is an example of the pipeline selecting a feature which shows strong discriminability under DBS, but which could be of muscular origin based on its edge-bound pattern. Session $\text{S10}_c$ also shows a significant improvement of the task performance under DBS ON, indicating that DBS does improve \handmotor function. Given this improvement, closing the loop based on electromyography (EMG)~\cite{Fleming2023, Wang2023} could be a viable option for this session. 
Looking at the epoch-wise power values (last row) for $\text{S10}_c$, we observe a wash-out effect in the OFF trials following a DBS ON block (starting from the second DBS ON block). These dynamics of a few minutes, mean = $431\,\text{s}\,(\pm\,39\,\text{\std})$, are in agreement with previously reported observations of an STN beta power rebound~\cite{BronteStewart2009} and a reappearance of symptoms~\cite{Temperli2003} after cessation of DBS.
\\
Session $\text{S12}_2$ (Figure~\ref{fig:marker_overview}, third column) is an example of a successful neural decoding, without a significant DBS effect on the CopyDraw score. The neural decoding was possible despite the limited cortical coverage (see white area of the topological plot), and even though the spectra do not modulate with DBS. This is an example of a non-controllable feature, which is therefore not suitable as a marker for aDBS. 
\\
Session $\text{S17}_3$ (Figure~\ref{fig:marker_overview}, last column) is finally an example of a commonly expected beta band attenuation, here visible in the ECoG signals. The feature is both informative and controllable, hence suitable for aDBS control.

\subsubsection{\textbf{Prototypical outcome types and their suitability for aDBS}}
\label{sec:discussion_scenarios}
Our approach for marker identification concatenates two decoding pipelines. As the first one provides the labels for the second one, it is interesting to analyze the linear dependency structure between the decoding performances of these two pipelines.
The correlation ($\text{r}=0.63$) between behavioral decoding (CopyDraw ROC AUC) and neural decoding correlation is visualized in Figure \ref{fig:outcome_types} A. It confirms that a good neural decoding---as expressed by a high neural decoding correlation---is more easily achieved if the CopyDraw ROC AUC is high. Therefore, the behavioral decoding alone already allows preselecting sessions/participants for which a neural decoding will more likely succeed. 
We use an intra-cluster correlation coefficient (ICC) to discriminate correlations caused by more linearly distributed values, from correlations caused by bimodal distributions.

Looking jointly at the CopyDraw scores and their neural predictions, we can distinguish six possible prototypical outcomes. Example sessions for each type are shown in Figure~\ref{fig:outcome_types} B - G. The six types are discussed in the following with a schematic summary provided in Figure~\ref{fig:outcome_types_schema}. 

Type 1: Behavior is dominantly modulated by DBS, and neural decoding is possible (high CopyDraw ROC AUC, high neural correlation, high ICC). Differences beyond the DBS effect are small, as seen by the large ICC. Since our experimental protocol used only a single DBS configuration (DBS ON vs. OFF), we cannot assess the impact of different DBS parameters, for instance, with reduced amplitudes. We hypothesize that the behavior would be modulated monotonically with intermediate DBS amplitudes, as previous studies have demonstrated amplitude-dependent modulation in MDS-UPDRS-III scores~\cite{Conovaloff2012}, behavioral tasks~\cite{Munoz2022} and electrophysiology~\cite{Anderson2020, Sinclair2019, Whitmer2012}. Based on this hypothesis, we consider such sessions suitable for investigating aDBS with a proportional modulation of the DBS amplitude~\cite{Arlotti2021}, which would exploit the potential intermediate values.
If the ICC is high, then DBS ON and OFF each result in a clear cluster, such that a single linear model may not be adequate to model the CopyDraw score. In this case, however, the (known) DBS label might already be predictive for the CopyDraw score.

Type 2: Behavior is modulated non-dominantly by DBS and neural decoding is possible (high CopyDraw ROC AUC, high neural correlation, low ICC). Beyond the DBS-induced difference in behavior, there are significant non-DBS related fluctuations in the CopyDraw score, which are captured by the neural decoding pipeline, i.e., a neural marker was found. Such sessions would be suitable for a threshold type aDBS~\cite{Busch2025}, since the limited \icc suggests that the non-DBS related fluctuations are already large compared to the DBS effect. Reduced DBS amplitudes would likely reduce the behavioral effect further.

Type 3: Behavior is modulated non-dominantly with DBS and neural decoding is not possible (high CopyDraw ROC AUC, low neural correlation, low ICC). In this case, DBS has an effect on behavior, which is not captured by the neural decoding pipeline. The low neural decoding could be caused by data quality issues, insufficient samples, or a decoding pipeline which is not suitable to capture the behavioral modulation. As \icc is also low, it is also possible that the DBS modulation is too subtle to be detectable with the given recording modalities.

Type 4: Only the neural decoding is significant. The participants \handmotor performance is not impacted by DBS, either because there are no deficits even under DBS OFF (ceiling effect, potentially due to micro-lesioning effect), or DBS with the particular configuration is not effective for addressing \handmotor deficits. In both cases, the neural marker is not suitable for aDBS with a focus on \handmotor symptoms.

Type 5: Behavior is dominantly modulated by DBS and neural decoding is not possible (high CopyDraw ROC AUC, low neural correlation, high ICC). In this case, the decoding pipeline should be investigated, as inefficiencies in the data quality or decoding pipeline could lead to a failed neural decoding. Improvements of the recording quality, or additional samples, could help to improve the neural decoding. We would expect a neural decoding of at least the DBS condition to be possible, as the behavior separates clearly (high ICC).

Type 6: Behavior is not modulated by DBS and no neural decoding is possible. In this case, the DBS configuration should be reconsidered. If general changes in symptom severity can be evoked by DBS, e.g., comparing MDS-UPDRS under DBS ON vs. OFF, then \handmotor capabilities might be within a ceiling effect with no significant improvement possible. If there is no observable change in general symptoms, then DBS might be insufficient, or other factors, such as the micro-lesioning effect, make an investigation with the proposed approach infeasible for such a session. 

Given the three metrics used to delineate the types above, we note that cases with low CopyDraw ROC AUC and high ICC are not possible, hence there are six types and not eight. It would be of interest to see if the presented types also reflect the populations in other studies. However, we are not aware of a study reporting behavioral modulation and neural correlates in such a way yet.

\subsection{Limitations and future work}
The following limitations of the present study warrant consideration and lead to implications for future work. 

First, a key confounder in the acute sessions is the micro-lesioning effect (MLE~\cite{Jech2012}), which lasts up to around 30 days post surgery~\cite{DeNeeling2025, Wu2025} and appears with high prevalence (e.g., 89\,\%)~\cite{Wu2025}. The MLE could potentially explain why no significant behavioral DBS effect was detected for a subset of our acute sessions. Symptom improvement comparable to clinical DBS can be caused by the MLE alone~\cite{Wu2025}. Therefore, \handmotor deficits might have been alleviated by the MLE, leaving little room for an additional DBS-induced improvement (ceiling effect). A recent study by Lee et al.~\cite{Lee2025}, disentangled the MLE and DBS effect on a finger tapping task by demonstrating that three weeks post surgery, DBS improves amplitude while an increased frequency is attributed to MLE. This shows that differences in motor behavior can still be induced by DBS despite the MLE, which could explain our observed successful \handmotor decoding, even for acute sessions. 

Second, all decoding pipelines were performed in a session-specific manner, and a prospective validation in which a marker identified in one session is used to drive or predict outcomes in a subsequent session was not performed. Addressing this topic will require a dedicated longitudinal study design, both to assess the cross-session stability of the biomarkers and to validate their predictive utility for aDBS. 

Third, each session compared only a single fixed DBS ON parameters set against the DBS OFF state. This binary contrast corresponds to an "extreme" case scenario, and future studies should assess how behavioral or neural markers respond to, e.g., intermediate stimulation amplitudes. 

Fourth, participants performed sessions under their standard medication regimen, and the timing of the dopaminergic medication was not systematically controlled. Given that levodopa independently modulates both motor performance~\cite{Alaei2025} and neural signals~\cite{Priori2004}, potentially acting on the same networks as DBS~\cite{Binns2025}, this constitutes a potential confounder. Future studies should consider fixed medication schedules or explicit medication-off conditions to disentangle DBS and dopaminergic effects on the identified markers.

Fifth, investigating the CopyDraw score, or the task performance, in relation to (long-term) MDS-UPDRS III outcomes would allow us to compare our outcomes to a broader range of studies. Despite being the most commonly used clinical outcome assessment~\cite{Mestre2024}, the MDS-UPDRS in its current form, is a topic of debate~\cite{Benesh2026, Daalen2025, Evers2019} due to various shortcomings.

Lastly, with prospective future work, we see a potential benefit in using the CopyDraw score or the task performance as proxies for the severity of symptoms, which can be used as optimization targets for automated programming of DBS parameters. Currently, automated DBS programming, promising a more efficient and comprehensive screening of parameters~\cite{Muller2025}, is primarily focused on (beta) band power features~\cite{Muller2025-review} from electrophysiological signals.
The advantage of considering band power features is the ability to read them out quickly. The disadvantage, on the other hand, is the indirect nature of the band power with respect to the severity of the symptoms, for which a correlation exists, but which might be different on group vs.~individual level~\cite{Gerster2025}. This is also reflected in the observation that even for aDBS, seven to eight programming sessions were necessary to find satisfactory aDBS settings~\cite{Busch2025}. The CopyDraw task could be an alternative with a slower readout which, however, is closer to actual \handmotor symptoms. Regarding success rates, we note that the prevalence of finding a CopyDraw effect is broadly comparable to the prevalence of finding power spectral peaks reported in a recent large cohort (ADAPT-PD)~\cite{Stanslaski2024}. Finally, the behavioral features of the CopyDraw task could allow at-home optimization of the DBS parameters, as it requires only common and inexpensive hardware (a notebook and a stylus), and would be feasible even without sensing-enabled devices. 

\section{Conclusion}

The presented data demonstrated that the CopyDraw task yields a robust success rate for identifying patient‑specific EEG and epidural ECoG neural markers that correlate with hand‑motor performance. Across six distinct outcome types, we formulate concrete recommendations for deploying these markers in the aDBS setting. The current results provide a quantitative framework and rationale for the design of future trials evaluating such marker‑based aDBS strategies.

\section{Methods}

\subsection{Patient population}
In total, 19 patients were included in two centers (17 at University of Freiburg Medical Center, Germany (UFMC) and 2 at Maastricht University Medical Center (MUMC), The Netherlands). Ethical approval was obtained by the local ethics boards (DRKS00028703 and NCT03079960, and NL84882.068.23). Participants provided their written consent prior to any study procedures. General demographic data are provided in Table~\ref{tab:demographics}.

\subsection{Recording and Stimulation settings}
\begin{figure}
    \centering
    \includegraphics[width=\linewidth]{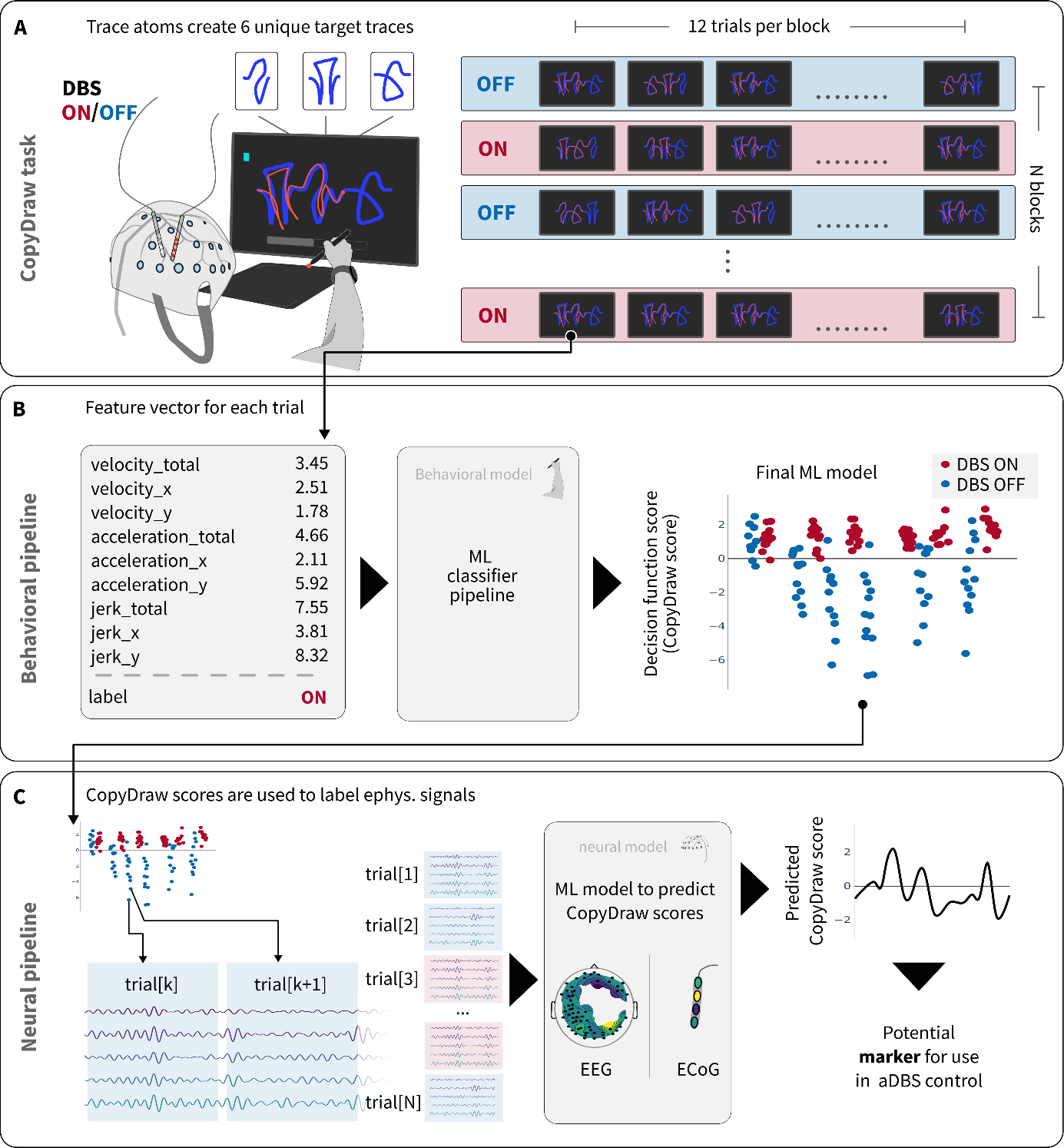}       
    \caption{The CopyDraw behavioral task and related analysis pipelines. \textbf{A} Participants use a stylus to control a drawing cursor on a screen. In each trial, the participant copies the target trace (blue) as accurately and quickly as possible. The inter-trial interval is self-paced. Twelve trials form a block. DBS stimulation is alternated between ON and OFF between blocks. \textbf{B} Behavioral pipeline: Behavioral features such as speed, acceleration, and jerk were extracted from the screen pixel coordinates (x, y) of the drawn trace. These features were used to train an LDA to classify DBS ON vs.~OFF states. The decision function value of the classifier was defined as a continuous CopyDraw score (one score per trial/trace). \textbf{C} Neural pipeline: On a supervised regression task, the CopyDraw scores were predicted from ECoG/EEG features. The output of such a regression model, i.e., the predicted scores, can be used as a potential control variable for an aDBS system. Abbreviations: DBS, deep brain stimulation; ECoG, electrocorticogram; EEG, electroencephalogram; ML, machine Learning; }
    \label{fig:copydraw_to_marker}
\end{figure}

Multiple blocks of the CopyDraw task~\cite{Castano2019} were performed with DBS (ON) and without DBS (OFF). Participants used a stylus and a tablet to copy a target trace displayed on a screen (Figure~\ref{fig:copydraw_to_marker} A). The participants were instructed to copy ``as quickly and as accurately as possible'', within a fixed time window of 6-10.5\,s (adjusted to the participants' capabilities). Each block consisted of 12 trials, separated by self-paced breaks. Details on the total number of trials conducted and the trial lengths can be found in the appendix Figure~\ref{fig:apx_trial_counts}.

\textbf{Cortical recordings} --
Electrophysiological signals were recorded from the cortex, using non-invasive high-density EEG and/or a four-channel transient epidural ECoG strip (Ad-Tech Medical Instrument Corporation, Oak Creek, USA) over the left hemisphere. ECoG strip electrodes were positioned during the surgical procedure of DBS electrode positioning. The angle of approach was defined naturally by the position of the burr hole, which was planned in such a way that it allowed for a safe trajectory to the STN, typically from the medial frontal gyrus. Placement was performed on the dominant hemisphere (left side) only (thus unilateral). Using the stereotactic frame and with the help of the “hand knob” sign~\cite{Yousry1997} the epidural path of the strip electrode was triangulated. The burr hole was epidurally extended like a trench for 1--2\,cm in the direction of the primary motor cortex (M1). The dura was dissected away from the bone and the strip electrode was inserted and advanced towards M1. The electrode positions are visualized in Figure~\ref{fig:apx_dbs_leads_and_ecog}.
At UFMC, EEG was recorded using BrainAmp DC amplifiers (Brain Products GmbH, Gilching, Germany) at different sampling rates between 1-5~kHz (fixed for each session), while ECoG was recorded via the Neuro Omega\textsuperscript{\texttrademark} at 22~kHz sampling rate using hardware filters between 0.07\,Hz and 10\,kHz. At MUMC, EEG was recorded using the Neuro Omega\textsuperscript{\texttrademark} (Alpha Omega Engineering, Nof HaGalil, Israel) at a sampling rate of 22\,kHz using hardware filters between 0.07~Hz and 10~kHz.

\textbf{Stimulation settings} --
 DBS was applied contralaterally (except for participant S8 and $\text{S1}_c$ see Appendix~\ref{apx:tbl_dbs_config}) to the hemisphere associated with the dominant hand, which was used to execute the CopyDraw task. At both centers, the Neuro Omega\textsuperscript{\texttrademark} was used to administer DBS. Clinicians evaluated bipolar DBS parameter configurations, taking into consideration side effects and the therapeutic windows. Bipolar configurations were chosen to reduce the stimulation artifact amplitude in the cortical recordings. This titration procedure defined the amplitude (mA), the stimulation channels, and the return channels. By default, symmetric stimulation pulses with an initial negative peak were applied at 130\,Hz with a 60\,$\mu$s pulse width. Settings were kept constant within each block, with DBS ON and OFF states alternating consecutively. Therapeutic parameters were used for the chronic sessions. The full DBS parameters are provided in appendix~\ref{apx:population}. 

\textbf{Data preprocessing} --
The preprocessing pipeline used mne~\cite{Gramfort2013mne} standard procedures for EEG and ECoG data. First, the data was resampled to $300\,\text{Hz}$, using the FFT-based method (mne.io.Raw.resample), which uses a truncation (brick-wall filter) at the new Nyquist frequency in the frequency domain to prevent aliasing. Inspection of the power spectral densities below 130~Hz showed aliasing stimulation artifacts for the MUMC EEG recordings (S18 and S19, EEG recording using the Neuro Omega\textsuperscript{\texttrademark}). Data for these sessions had the stimulation artifact removed by linear interpolation of the signal from $-30\mu s$ to $360\mu s$ relative to the first negative peak of each stimulation pulse in the raw 22~kHz data, before resampling. No aliasing artifacts were visible after interpolation or for the other sessions.
All intermediate data objects were persisted at 300\,Hz as this allowed to validate the presence of the DBS stimulation artifact at 130\,Hz. Raw data was plotted and bad channels were visually excluded. Next, for EEG data only, an independent component analysis (ICA) was fitted on data which was further down sampled to 200\,Hz (for algorithmic efficiency) and frequency filtered to (1--30)\,Hz. Artifactual ICA components were identified by visual inspection and excluded from the broad band raw data (at 300\,Hz). Further processing within the filterbank SPoC/CSP pipelines was conducted in source (ICA) space for computational efficiency. Sessions with ECoG data used a common average re-referencing, while no referencing was applied to EEG.

\textbf{CopyDraw task} --
The CopyDraw task~\cite{Castano2019} is designed to capture \handmotor features under different DBS stimulation conditions. A schematic overview of the task and the derivation of the CopyDraw score are provided in Figure~\ref{fig:copydraw_to_marker} (panels A, B). During the task, participants use a stylus and tablet (\textit{One by Wacom\textregistered}, Wacom, Kazo, Japan) to copy a target trace, by controlling the cursor on a screen in a comfortable viewing distance ($\approx$ 70\,cm). The target traces are combinations of three trace atoms, i.e., characters, which on purpose avoid resemblance to common characters of the Latin alphabet. Participants are requested to copy-draw the template \textit{as quickly and as accurately as possible}, within a given time window. The time window is adjusted to the participants' capabilities and ranged from 6 to 10.5\,s (see Figure~\ref{fig:apx_trial_counts} C), such that the participant would usually make it through more than 2/3 of the trace, but would be rushed to finish in time. Copy-drawing a single trace is referred to as a trial. The inter-trial intervals are self-paced and the participant has to express readiness by moving the cursor over a cyan square in the upper left of the screen before moving to the beginning of the trace. Reaching the beginning of the trace activates the drawing and the time-out mechanism. The time left within the trial is indicated by a bar on the bottom of the screen. We combined 12 trials into a block and offered the possibility for breaks between each block. DBS was alternated for each of the blocks (DBS OFF, ON, OFF, ON, ...). The 35 sessions comprise a total of 5292 CopyDraw trials with mean (standard deviation) trial counts per session of 73 ($\pm\,18$) for DBS ON and 72 ($\pm\,18$) for DBS OFF. A detailed overview of trial counts and exclusions due to the lab protocol or for interrupted drawings (217 of 5292) is provided in Figure~\ref{fig:apx_trial_counts} A and B. Prior to the first recorded block, participants completed a familiarization phase to ensure they understood the task and could reliably control the cursor with the stylus. 

\begin{table}
\caption{Participant demographics - \textit{UPDRS on} and \textit{UPDRS off} refer to the MDS-UPDRS part III under medication on and off condition. All values are as of the regular inpatient visit closest, but prior to the experiment session. For participants with externalized leads (S1-S7, S12, S14-S19) all values correspond to pre-surgery tests. Three UPDRS scores were not available from the medical records and are marked as NA. \textit{Symptom side} is the body hemisphere, at which the initial motor symptoms were observed. Abbreviations: \textit{f}, female; \textit{m}, male; \textit{l}, left; \textit{r}, right; \textit{LEDD}, levodopa equivalent daily dose;  \textit{UPDRS}, Unified Parkinson's Disease Rating Scale;  }
\label{tab:demographics}
\begin{tabular}{lrrlllllr}
\toprule
Subject & Age [years]& LEDD& Sex & Disease & Symptom & UPDRS & UPDRS & Handedness\\
  &  &    &   & duration [years]& side & on & off & \\
\midrule
S1 & 55 & 1240 & m & 5 & l & 25 & NA & l\\
S2 & 54 & 885 & f & 15 & r & NA & 18 & r\\
S3 & 56 & 1190 & m & 11 & r & 20 & 41 & r\\
S4 & 44 & 1775 & m & 4 & r & 15 & 23 & r\\
S5 & 69 & 2040 & m & 18 & r & 15 & 42 & r\\
S6 & 63 & 698 & m & 15 & l & 35 & 69 & r\\
S7 & 55 & 1240 & m & 7 & l & 13 & 34 & r\\
S8 & 63 & 830 & m & 15 & l & 17 & 22 & r\\
S9 & 61 & 210 & m & 1 & r & 13 & 27 & r\\
S10 & 51 & 1095 & f & 16 & bilateral & 15 & 26 & r\\
S11 & 60 & 493 & m & 8 & l & 20 & 14 & r\\
S12 & 64 & 1775 & f & 15 & r & 31 & 68 & r\\
S13 & 71 & 1084 & m & 13 & l & 38 & 53 & r\\
S14 & 64 & 1325 & f & 9 & l & 22 & 58 & r\\
S15 & 52 & 1669 & f & 16 & l & 28 & 44 & r\\
S16 & 58 & 1197 & f & 8 & r & 24 & 15 & r\\
S17 & 68 & 809 & f & 12 & l & 36 & NA & r\\
S18 & 64 & 1525 & f & 5 & r & 14 & 54 & r\\
S19 & 61 & 1800 & m & 5 & r & 15 & 37 & r \\ 
\bottomrule
\end{tabular}
\end{table}

\subsection{Data analysis pipeline for behavioral decoding}
\label{sec:behavioral_pipeline}
\subsubsection{Computing CopyDraw scores}

The \handmotor behavior of patients during the CopyDraw task was translated into continuous scalar labels, herein referred to as CopyDraw scores (one score per trial). To derive these scores, first, a nine-dimensional feature vector $\mathbf{v}\in\mathbb{R}^{9}$ was extracted (Figure~\ref{fig:copydraw_to_marker} B) for each trial. The feature vector contained mean absolute values for speed, acceleration, and jerk in the x- and y-directions and in magnitudes. Following the procedure outlined by \Castano et al.~\cite{Castano2019}, these metrics were computed as the first, second, and third derivatives with respect to time of the drawn trace, respectively.
The simplified nine-dimensional feature set was chosen over the more complex 36-dimensional angularly binned feature set initially proposed by \Castano et al.~\cite{Castano2019}, because the larger feature set did not lead to a significant improvement in the decoding of the DBS condition at the group level (Figure~\ref{fig:apx_feature_set_session_type_signal_modality} A). We chose the smaller feature set to improve robustness of the decoding pipeline that is introduced in the following paragraph.

The features were clipped to $\pm\,3$ standard deviations from the mean within the training data to constrain outliers. After standard-scaling (z-scaling), the features were used as input to a linear discriminant analysis (LDA) classifier using Ledoit-Wolf~\cite{Ledoit2004} regularization, which was trained to classify the DBS condition (ON vs.~OFF). Per trial, the corresponding LDA decision function value determined the continuous \textit{CopyDraw score}. A CopyDraw score $\geq0$ was associated with DBS ON and a CopyDraw score $<0$ was associated with DBS OFF. The pipeline is a modification of the one proposed by \Castano et al.~\cite{Castano2020}, removing the linear de-trending and changing the outlier removal to simple clipping, to prevent leakage of information across the cross-validation scheme introduced in section~\ref{sec:eval_and_cv} (details of the processing pipeline and changes are visualized in Figure~\ref{fig:apx_adjusted_behavioral_pipeline}).
Shapley additive explanation (SHAP) values~\cite{Lundberg2017} were computed to post-hoc attribute the importance of individual behavioral features.

\subsubsection{Computing task performance}
Because of the aforementioned DBS ON vs.~OFF classification task, the CopyDraw score quantifies whether a drawn trace looks more like it was drawn under DBS ON or OFF. It does not necessarily reflect the overall accuracy of the copied trace with respect to the target template, or the speed of copying. For this purpose, we introduced another metric, the \textit{task performance}. Using dynamic time warping (DTW)~\cite{Giorgino2009}, samples of the drawn trace were matched to the template~\cite{Castano2019}, with the mean of the euclidean distances ($\|\delta\|_{trace}=\frac{1}{n_{trace}}\sum_i^{n_{trace}}\delta^{i\rightarrow j}_i$) between each trace sample $i\in\{1,2,...,n_{trace}\}$ and the matching template points $j\in\{1,2,...,n_c\}$ quantifying the accuracy. Lower values indicate a higher accuracy. To quantify the extent of the template the participant was able to trace, the last matched point on the template ($n_c$) relative to the total number of template points ($n_{total}$) was considered, defining the task performance as 
\begin{equation}
\text{task performance} = \frac{n_c/n_{total}}{\|\delta\|_{trace}}\quad.
\end{equation}

This fraction reflects the trade-off between speed and accuracy in a non-linear relation, which would not be captured by a linear model such as LDA alone. 
Similarly to the LDA fit leading to the CopyDraw score, we trained an LDA using the task performance as a single feature for the classification of DBS ON and OFF. 
The area under the receiver operating characteristic curve (ROC AUC) of both classifiers was used to compare the information content about the DBS state in the CopyDraw features with the one in the task performance.

\subsection{Data analysis pipeline for neural decoding}
\subsubsection{Predicting behavioral scores from neural signals}
For both types of behavioral labels, provided by the task performance or by the CopyDraw scores, the same neural decoding pipelines were used.\\
A supervised regression pipeline was used to predict the scores from neural signals obtained from EEG or ECoG (Figure~\ref{fig:copydraw_to_marker} C). For the EEG data, our pipeline was motivated by \Castano et al.~\cite{Castano2020}, but instead of using a single band source-power comodulation (SPoC)~\cite{Dahne2014}, we applied a filter bank SPoC (FBSPoC)~\cite{Castano2017} followed by a linear regression with quadratic regularization (ridge regression, $\alpha=1$). This change was applied to simplify the number of hyperparameters and to allow multiband feature combinations. For each of the frequency bands $\{\theta$=(4--8)\,Hz, $\alpha$=(8--12)\,Hz, $\beta$=(12--30)\,Hz, $\gamma$=(30--45)\,Hz, $\gamma_{\text{high}}$=(55--90)\,Hz$\}$, eight SPoC filters were extracted. We limited the frequency ranges to up to 90\,Hz to stay clear of the second harmonic of the line noise (50\,Hz) as well as the DBS artifact at 130\,Hz.
Out of all 40 filters (5 freq. bands times 8 SPoC filters), eight were selected using a minimum redundancy maximum relevance (MRMR)~\cite{Peng2005} feature selection. A detailed overview of the differences in the decoding pipeline between \Castano et al.\cite{Castano2020} and ours is provided in Figure~\ref{fig:apx_adjusted_neural_pipeline} in the appendix. For the ECoG data, band power features were extracted for each of the four ECoG channels using the same frequency bands as those used for EEG. MRMR selection was applied to reduce the input feature dimension to eight features used as input for a final linear regression (ridge regression, $\alpha=1$). The main purpose of this selection is to keep the EEG and ECoG pipelines as similar as possible.
The output of this decoding step consists of a neural marker characterized by frequency and spatial filters.

\subsubsection{Analyzing the controllability of decoded neural markers}
A neural marker is only suitable for aDBS if it can be controlled by DBS. To evaluate the impact of DBS on an individual neural marker, we analyzed how well its features can predict the DBS condition (ON/OFF). For this purpose, we kept the frequency and spatial filters determined by the regression pipelines mentioned above fixed, but trained and evaluated an LDA classifier instead of the ridge regression in the final step. We considered features which can be used to predict the DBS condition as \textit{controllable} by DBS. 

To visualize individual features, we explored patterns~\cite{Haufe2014} of the spatial filters, the power spectral densities (PSD) of the spatially filtered signals, as well as the mean band power of the spatially and band filtered signal per trial. The PSDs are calculated using the multitaper~\cite{Slepian1978} approach implemented in mne~\cite{Gramfort2013mne} (version 1.11.0) using default parameters.

\subsection{\textbf{Pipeline evaluation metrics and cross‑validation}}
\label{sec:eval_and_cv}
Both the behavioral and neural pipelines were evaluated using a chronological cross-validation (chrono-CV) scheme, which paired chronologically adjacent blocks of different DBS conditions. As a consequence, each test fold contained exactly one DBS ON block and one OFF block, while all other data were used for training. Folds were created such that each block was part of one testing set, after removing blocks with fewer than six valid trials (i.e., $<$ 50\,\% of a standard block). In case of an odd number of available blocks, the first block was omitted for pairing up blocks in the test set, but was considered in the training data. This was a deliberate choice to best avoid a potentially incomplete familiarization. The chrono-CV approach is a trade-off between respecting the impact of slow non-stationarities and being data efficient with regard to the available data for training and evaluation.\\
To evaluate the classification of the DBS condition based on behavioral features (behavioral pipeline, see section~\ref{sec:behavioral_pipeline}), the ROC AUC was computed for each test fold. The decoding performance is reported as the mean ROC AUC across all chrono-CV folds, hereafter referred to as \textit{CopyDraw ROC AUC}. The CopyDraw scores used as ground-truth labels for downstream neural decoding were not extracted from these individual CV folds. Instead, they were generated using a behavioral pipeline fitted to the entire behavioral dataset, ensuring a mathematically uniform projection of the behavioral features. 
The neural decoding performances were measured by Pearson's correlation between the predicted scores and the actual scores (CopyDraw score or task performance) of the trials in the test fold. We report the mean correlation across chrono-CV as \textit{\neuralscore}. The neural classification of the DBS condition uses the same chrono-CV and is again reported using the ROC AUC metric.

\subsection{Statistical analyses and data availability}
Chance levels were estimated using the 95\,\% percentile of a permutation distribution~\cite{Ojala2010permutation} with n=1000 shuffles of the training labels. Significance was assessed against a Type-I error sensitivity of $\alpha=0.05$. Welch's t-tests were used for their robustness in small samples~\cite{Dewinter2013, Delacre2017} whenever we had no indication of a violation of the normality assumption. Normality was tested using Shapiro-Wilk tests. Non-parametric Mann-Whitney U-tests were used for distributions that violated the normality assumption. Linear regressions were fitted using the ordinary least squares approach (OLS) and regression performance is reported as Pearson's correlation (r), $P$ values of the correlation, and explained variance ($\text{R}^2$). As the CopyDraw scores are labels from two distinct DBS conditions, the intracluster correlation coefficient (ICC), \mbox{\icc $:= \sigma_b^2/(\sigma_b^2 + \sigma_w^2)$}, was considered to quantify the bimodal nature of the CopyDraw score distributions. Here $\sigma_b$ and $\sigma_w$ are the between- and within-cluster standard deviations, respectively. For every multiple comparisons problem, cluster-based permutation tests were applied~\cite{Maris2007}.

\subsection{Data availability}
The analysis was performed using custom Python scripts utilizing algorithms from scipy~\cite{Scipy2020}, statsmodels~\cite{Seabold2010} and mne~\cite{Gramfort2013mne}. The full data set and analysis scripts will be publicly available via the Radboud Data Repository at https://doi.org/10.34973/dxt6-qn46\,.

\bibliographystyle{vancouver}   
\bibliography{bibliography}

\section{Acknowledgments}

MD receives funding from the Dareplane collaboration project, which is co-funded by PPP Allowance awarded by Health~Holland, Top Sector Life Sciences \& Health, to stimulate public-private partnerships as well as by a contribution from the Dutch Brain Foundation. The PD-Interaktiv I study is funded by the Bundesministerium für Bildung und Forschung (Grant 16SV8011). 
VAC receives a collaborative grant from BrainLab (Munich, Germany). He serves as an advisor for Aleva (Lausanne, Switzerland), Ceregate (Hamburg, Germany), Cortec (Freiburg, Germany) and Inbrain (Barcelona, Spain). He has an ongoing IIT with Boston Scientific (USA). He has received travel support and honoraria for lectures from Boston Scientific (USA), UNEEG Medical (Munich, Germany), and Precisis (Heidelberg, Germany). The authors thank the EuroDBS Alliance (EuroDBS.org) for feedback and discussions on this work. 
BS is funded by the Berta-Ottenstein-Programme for Advanced Clinician Scientists, Faculty of Medicine, University of Freiburg. He serves as a scientific advisor for Precisis (Heidelberg, Germany) and has received a research grant from Ceregate (Hamburg, Germany) and travel support from Boston Scientific (Marlborough, MA, USA) and honoraria as a speaker and travel support from Medtronic PLC (Dublin, Ireland) and Insightec (Tirat Carmel, Israel).
PR has received research support from the Else Kröner Fresenius Foundation, Fraunhofer Foundation (ATTRACT), German Ministry for Economic Affairs and Energy, and Medical Faculty of the University of Freiburg. He has received personal honoraria for lectures or advice from Boston Scientific, Brainlab, Inomed, and Fraunhofer Foundation and is a consultant to Boston Scientific, Brainlab, and Inomed. JP is funded by the Deutsche Forschungsgemeinschaft (DFG) under the Walter Benjamin Program (Project number 510112977). The other authors declare not to have conflicting interests.

\section{Author Contributions}
All authors: Writing - Review and Editing. MD: Conceptualization, Methodology, Software, Validation, Formal Analysis, Investigation, Data Curation, Writing - Original Draft, Visualization, Project Administration. VAC: Conceptualization, Methodology, Resources, Supervision, Project Administration, Funding Acquisition. BS: Investigation, Project Administration. PR: Investigation. TP: Investigation. MR: Visualization. SG: Investigation, Data Curation. YT: Resources, Supervision, Funding Acquisition. MLFJ: Conceptualization, Investigation, Resources, Writing - Original Draft, Supervision, Project Administration, Funding Acquisition. MT: Conceptualization, Methodology, Investigation, Resources, Writing - Original Draft, Supervision, Project Administration, Funding Acquisition. JP: Conceptualization, Methodology, Software, Investigation, Writing - Original Draft, Supervision, Project Administration, Funding Acquisition.

\appendix
\setcounter{figure}{0}
\renewcommand{\thefigure}{S\arabic{figure}}
\setcounter{table}{0}
\renewcommand{\thetable}{S\arabic{table}}

\section{Appendix}
\label{apx:population}
\begin{table}[ht]
\centering
\scriptsize 
\setlength{\tabcolsep}{3pt} 

\caption{DBS parameters and neural recording modality. If not remarked otherwise, stimulation was provided with 130\,Hz, 60\,$\mu$s, initial negative pulses. Symmetric pulse shapes were used for all acute sessions (Session $\neq$ c), and with implanted pulse generator defaults for the chronic sessions.}

\label{apx:tbl_dbs_config}

\begin{tabular}{
    p{1.0cm} 
    p{0.8cm} 
    p{1.5cm} 
    p{1.0cm} 
    p{1.5cm} 
    p{1.5cm} 
    p{2.0cm} 
    p{2.0cm} 
    p{1.2cm} 
}
\toprule

\textbf{Subject} & 
\textbf{Session} & 
\textbf{Amp. [mA]} & 
\textbf{Lat.} & 
\textbf{Stimulation} & 
\textbf{Return} & 
\textbf{Electrode} & 
\textbf{Remark} & 
\textbf{Modality} \\

 & & & & \textbf{contacts} & \textbf{contacts} & & & \\
\midrule

S1 & 2 & 3.00 & R & 8 & 5,6,7 & Boston Vercise™ & & EEG \\
S1 & 3 & 3.00 & R & 5,6,7 & 8 & Boston Vercise™ & & EEG \\
S1 & c & 2.40 & & 8 & 5,6,7 & Boston Vercise™ & bilateral & EEG \\
\midrule

S2 & 2 & 6.00 & L & 8 & 5,6,7 & Boston Vercise™ & & EEG \\
S2 & 3 & 6.00 & L & 8 & 5,6,7 & Boston Vercise™ & & EEG \\
S2 & c & 4.50 & L & 8 & 5,6,7 & Boston Vercise™ & & EEG \\
\midrule

S3 & 2 & 7.00 & L & 8 & 5,6,7 & Boston Vercise™ & & EEG \\
S3 & 3 & 7.00 & L & 8 & 5,6,7 & Boston Vercise™ & & EEG \\
S3 & c & 6.00 & L & 8 & 5,6,7 & Boston Vercise™ & & EEG \\
\midrule

S4 & 2 & 5.00 & L & 8 & 5,6,7 & Boston Vercise™ & & EEG \\
S4 & 3 & 5.00 & L & 8 & 5,6,7 & Boston Vercise™ & & EEG \\
S4 & c & 3.50 & L & 8 & 5,6,7 & Boston Vercise™ & & EEG \\
\midrule

S5 & 2 & 6.50 & L & 8 & 5,6,7 & Boston Vercise™ & & EEG \\
S5 & 3 & 6.50 & L & 8 & 5,6,7 & Boston Vercise™ & & EEG \\
S5 & c & 6.50 & L & 8 & 5,6,7 & Boston Vercise™ & & EEG \\
\midrule

S6 & 1 & 3.50 & L & 8 & 5,6,7 & Boston Vercise™ & & EEG \\
S6 & 2 & 3.00 & L & 8 & 5,6,7 & Boston Vercise™ & & EEG \\
S6 & c & 3.50 & L & 8 & 5,6,7 & Boston Vercise™ & & EEG \\
\midrule

S7 & 2 & 4.50 & L & 8 & 5,6,7 & Boston Vercise™  & & EEG \\
S7 & 3 & 5.00 & L & 8 & 5,6,7 & Boston Vercise™  & & EEG \\
S7 & c & 3.50 & L & 8 & 5,6,7 & Boston Vercise™  & & EEG \\
\midrule

S8 & c & 5(Volt) & & 2 & 3 & Medtronic 3389 & bilateral, 30 $\mu$s & EEG \\
\midrule

S9 & c & 2.60 & L & 8 & 5,6,7 & Boston Vercise™ & & EEG \\
\midrule

S10 & c & 3.00 & L & 8 & 6 & Boston Vercise™ & asymmetric& EEG \\
\midrule

S11 & c & 4.50 & L & 8 & 7 & Boston Vercise™ & asymmetric, $40\,\mu s$ & EEG \\  
\midrule

S12 & 2 & 6.50 & L & 1 & 2,3,4 & Boston Vercise™ & & EEG \\
S12 & 4 & 6.50 & L & 1 & 2,3,4 & Boston Vercise™ & & EEG \\
\midrule

S13 & c & 2.00 & L & 5,6,7 & 8 & Boston Vercise™ & & EEG \\
\midrule

S14 & 2 & 6.00 & L & 2,3,4 & 5,6,7 & Boston Vercise™ & & ECoG \\
S14 & 4 & 5.00 & L & 2,3,4 & 5,6,7 & Boston Vercise™ & & ECoG \\
\midrule

S15 & 2 & 5.00 & L & 2,3,4 & 5,6,7 & Boston Vercise™ & & ECoG \\
S15 & 4 & 5.00 & L & 2,3,4 & 5,6,7 & Boston Vercise™ & & ECoG, EEG \\
\midrule

S16 & 2 & 3.50 & L & 2,3,4 & 5,6,7 & Boston Vercise™ & & ECoG \\
S16 & 4 & 4.50 & L & 2,3,4 & 5,6,7 & Boston Vercise™ & & ECoG \\
\midrule

S17 & 2 & 4.00 & L & 5,6,7 & 2,3,4 & Boston Vercise™ & & ECoG \\
S17 & 3 & 4.00 & L & 5,6,7 & 2,3,4 & Boston Vercise™ & & ECoG \\
\midrule

S18 & 3 & 3.00 & L & 8 & 5,6,7 & Medtronic \mbox{SenSight™} & & EEG \\
\midrule

S19 & 2 & 3.00 & L & 2,3,4 & 5,6,7 & Medtronic \mbox{SenSight™} & & EEG \\

\bottomrule
\end{tabular}
\end{table}

\begin{figure}
    \centering
    \includegraphics[width=\linewidth]{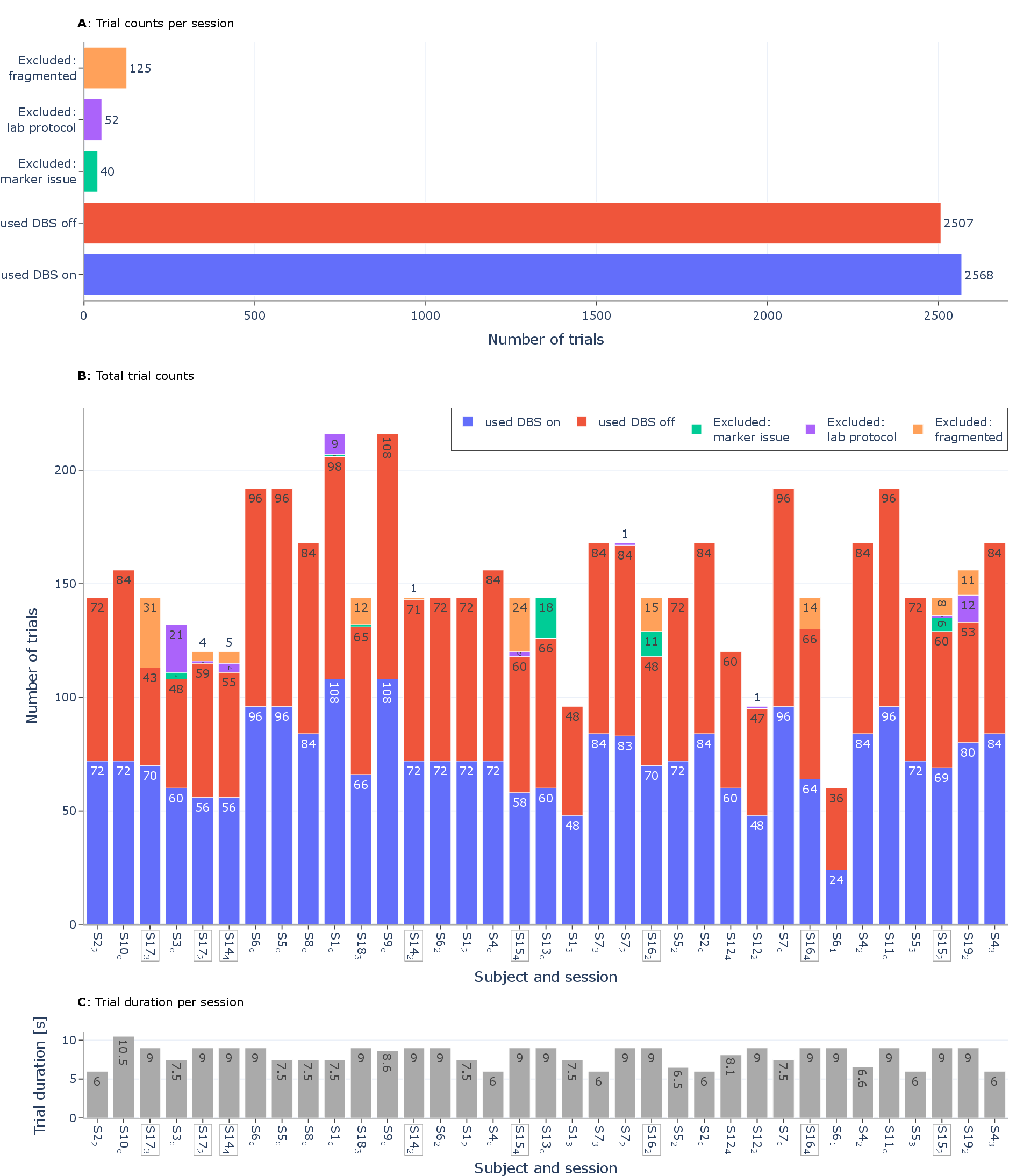}      
    \caption{Trial counts and duration. \textbf{A} shows total trial counts for each DBS condition as well as for the three exclusion criteria: \textit{Excluded: marker issues} - issues with synchronization, resulting in missing markers in the EEG or ECoG data. \textit{Excluded: lab protocol} - any exclusion due to reasons mentioned in the lab protocol, e.g., participant not adhering to the task, or falling asleep. \textit{Excluded: fragmented} - if the stylus lost connection during the execution of a trial. Note that the fragmentation was not tracked for trials of the study NCT03079960, hence no trials were removed for this reason from S1 to S13, see B. \textbf{B} shows the trial counts per session, again colored according to the DBS or exclusion criterion, using the same color codes as in A. The sessions are sorted according to the mean CopyDraw ROC AUC (see Figure~\ref{fig:behav_overview} A). Tick labels with a frame correspond to sessions with invasive ECoG data. \textbf{C} shows the trial durations for each session in seconds. Sessions are sorted as in B.}
    \label{fig:apx_trial_counts}
\end{figure}

\begin{figure}
    \centering
    \includegraphics[width=\linewidth]{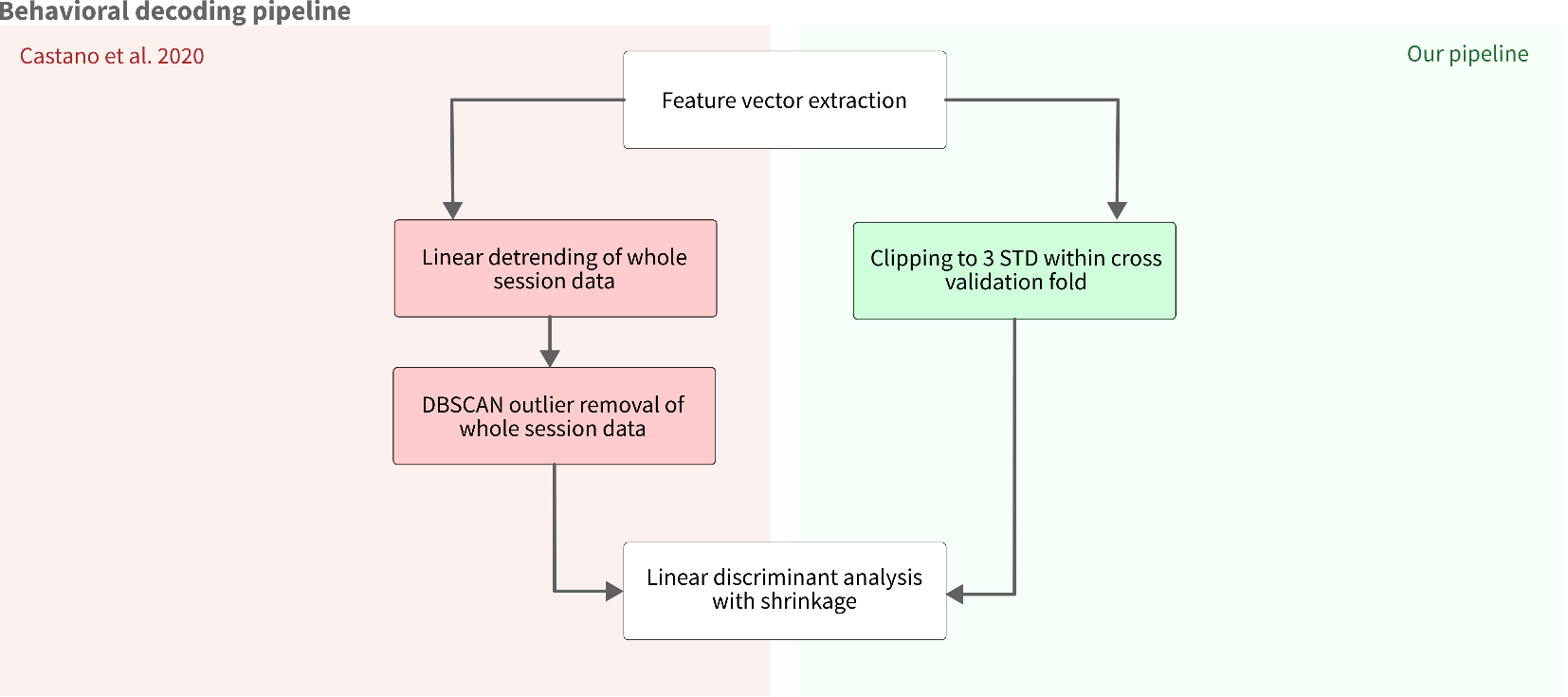}      
    \caption{Behavioral pipeline, \Castano et al. 2020~\cite{Castano2020} and ours. We replaced the linear de-trending and DBSCAN~\cite{Ester1996} outlier removal, which used the full data set, with a simple clipping of feature values to three standard deviations (STD) estimated from the training data within each cross-validation fold. The adjustment was made to generate a more robust estimation of the generalization performance due to the cross-validation scheme.}
    \label{fig:apx_adjusted_behavioral_pipeline}
\end{figure}

\begin{figure}
    \centering
    \includegraphics[width=\linewidth]{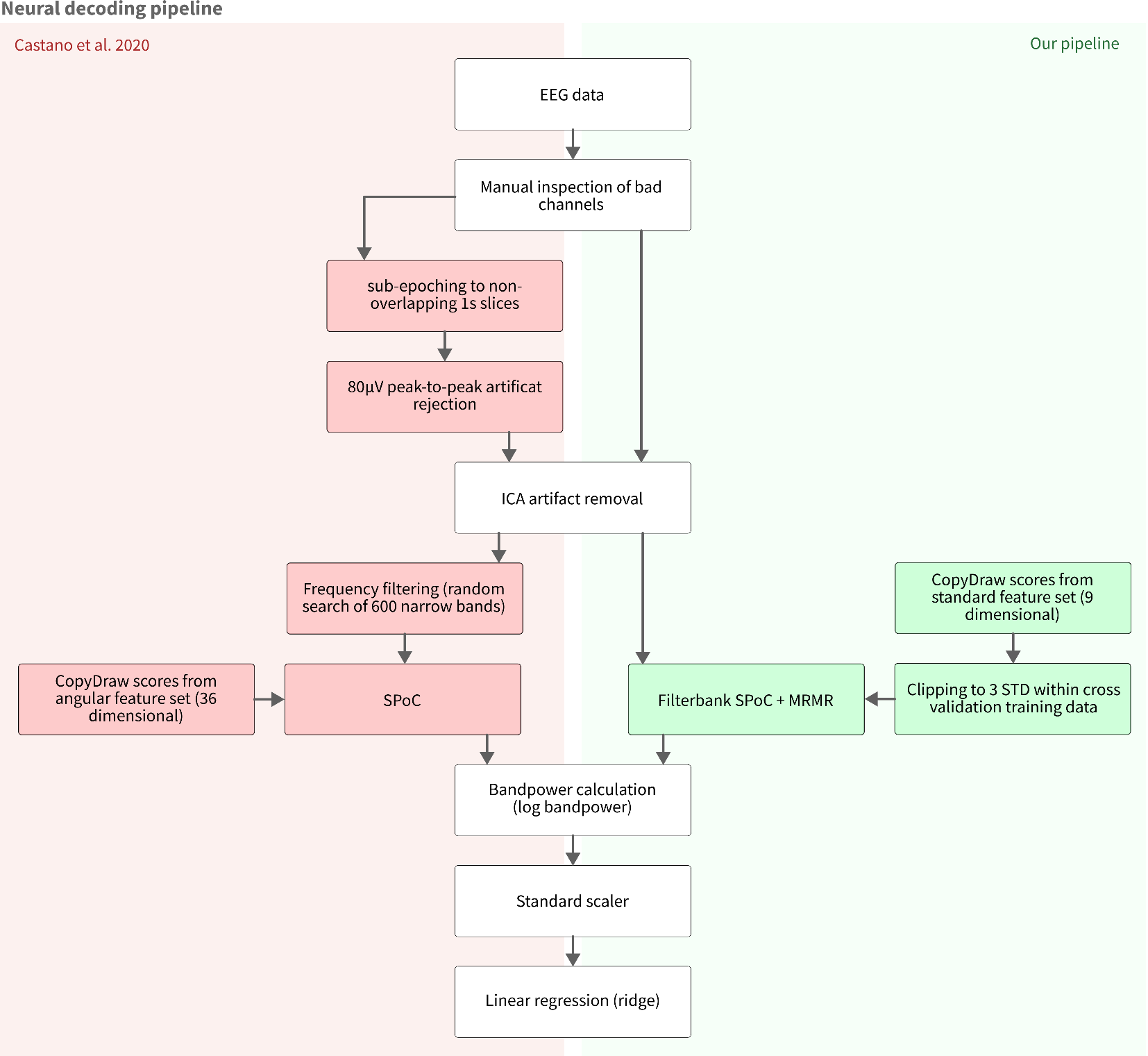}      
    \caption{Neural pipeline, \Castano et al. 2020~\cite{Castano2020} and ours. First, we removed the sub-epoching and subsequent removal based on peak-to-peak voltage changes of $80\mu V$, to simplify the pipeline. Next, we replaced the single band source power comodulation (SPoC)~\cite{Dahne2014} with a filterbank SPoC (FBSPoC) approach with five fixed frequency bands (\{$\theta$=(4--8)\,Hz, $\alpha$=(8 - 12)\,Hz, $\beta=$(12--30)\,Hz, $\gamma$=(30 - 45)\,Hz, $\gamma_{\text{high}}$=(55 - 90)\,Hz\}). A minimum redundancy maximum relevance (MRMR)~\cite{Peng2005} feature selection approach is used to select eight SPoC filters in total. For a proper cross-validation, the pipeline by \Castano et al. would require an additional split of the training data to provide a validation fold for optimizing the frequency band parameters. With the FBSPoC approach, we circumvent this selection and have it as part of the pipeline fitting, while also allowing multiband features to be considered for the final regression. The behavioral labels used to fit our pipelines are different as well. \Castano et al. 2020 used CopyDraw scores from the full 36 dimensional feature vectors (\textit{angular} feature set), while we use CopyDraw scores derived from the nine dimensional feature vectors (\textit{standard} feature set). We chose to report neural decoding using the standard feature set as no significant cross-session difference was observed for the behavioral decoding (Figure~\ref{fig:apx_feature_set_session_type_signal_modality}), hence the smaller feature set would be the more robust choice in light of our limited data ($\approx$100--150 trials per session). The scores derived from the smaller dimensional features should lead to more robust results, less susceptible to overfit.}
    \label{fig:apx_adjusted_neural_pipeline}
\end{figure}

\newpage

\begin{figure}
    \centering
    \includegraphics[width=\linewidth]{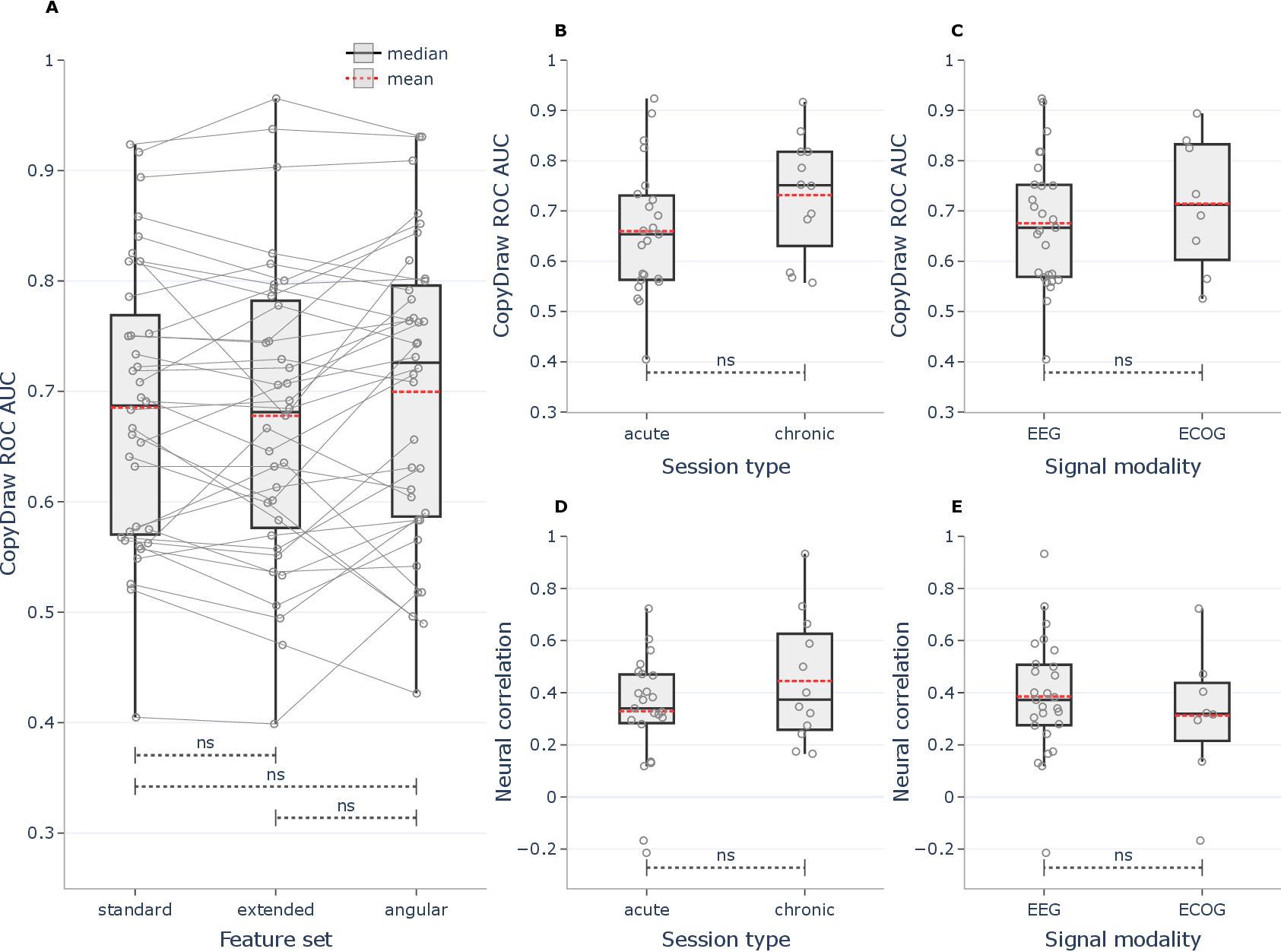}      
    \caption{CopyDraw ROC AUC and neural correlation of subsets. \textbf{A} shows the CopyDraw ROC AUC for different feature sets. The \textit{standard} $\in\mathbb{R}^{9}$ feature set contained speed, acceleration, and jerk in x- and y-directions and as magnitude values. The \textit{extended} $\in\mathbb{R}^{16}$ feature set uses the \textit{standard} set and distance features from dynamic time warping as well as the pre-trial time, which is the time between signaling readiness by moving to the cyan square and moving to the template start. The \textit{angular} $\in\mathbb{R}^{36}$ feature set contains the \textit{standard} feature set and angularly binned versions of these features according to \Castano et al.~\cite{Castano2019}. While the angular feature set has the highest mean and median ROC AUC, none of the differences are significant (Shapiro-Wilk: $P_{standard}=0.37$, $P_{extended}=0.96$, $P_{angular}=0.94$, Welch's t-tests $P_{standard,extended}=0.86$, $P_{standard,angular}=0.45$, $P_{extended, angular}=0.36$). \textbf{B} shows the CopyDraw ROC AUC for chronic and acute sessions. No significant difference (Shapiro-Wilk: $P_{chronic}=0.48$, $P_{acute}=0.59$, Welch's t-test $P=0.11$). \textbf{C} shows the CopyDraw ROC AUC for sessions with EEG recording and sessions with ECoG recordings. No significant difference  (Shapiro-Wilk: $P_{EEG}=0.55$, $P_{ECOG}=0.74$, Welch's t-test $P=0.48$). \textbf{D} shows the neural correlation for chronic and acute sessions. No significant difference  (Shapiro-Wilk: $P_{chronic}=0.39$, $P_{acute}=0.09$, Welch's t-test $P=0.17$). \textbf{E} shows the neural correlation for sessions with EEG recording and sessions with ECoG recordings. No significant difference (Shapiro-Wilk: $P_{EEG}=0.71$, $P_{ECOG}=0.72$, Welch's t-test $P=0.48$).} 
    \label{fig:apx_feature_set_session_type_signal_modality}
\end{figure}

\begin{figure}
    \centering
    \includegraphics[width=\linewidth]{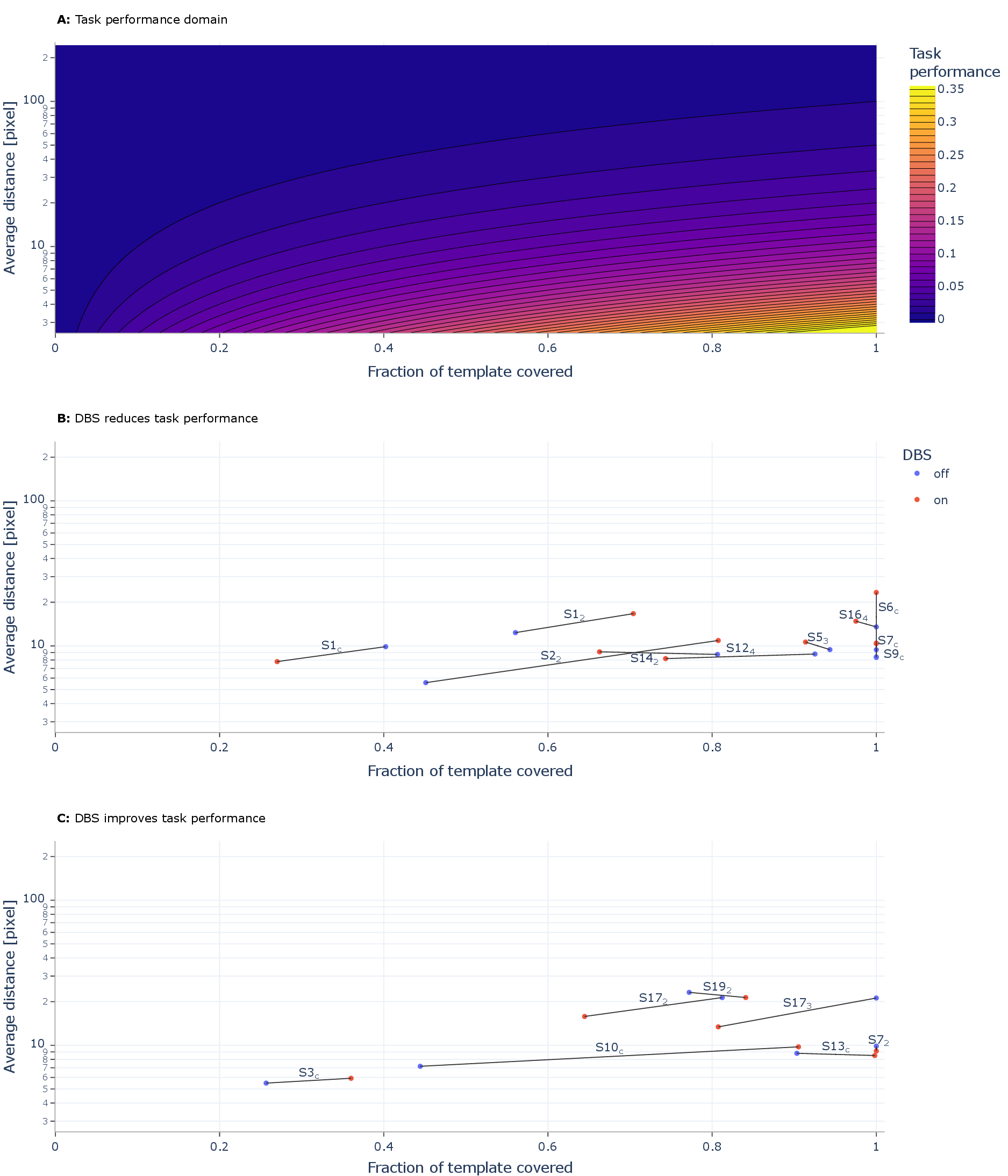}      
    \caption{Task performance as trade-off between speed and accuracy. \textbf{A}: Contour map of the task performance. The x-axis represents the amount of the template that has been matched by the drawn trace (matched using dynamic time warping). The y-axis represents the average distance between the drawn trace and the template on a log10 scale. The range of the y-axis is calibrated to contain all extreme values observed in trials from sessions with a significant DBS ON vs. OFF effect. \textbf{B}: Scatter plot of median values under DBS ON (red) and OFF (blue) for sessions with a significant decrease in task performance induced by DBS. Axes are the same as in panel A. \textbf{C}: Same as B but for sessions with a significant increase in task performance. }
    \label{fig:apx_task_perf_speed_vs_accuracy}
\end{figure}

\begin{figure}
    \centering
    \includegraphics[width=\linewidth]{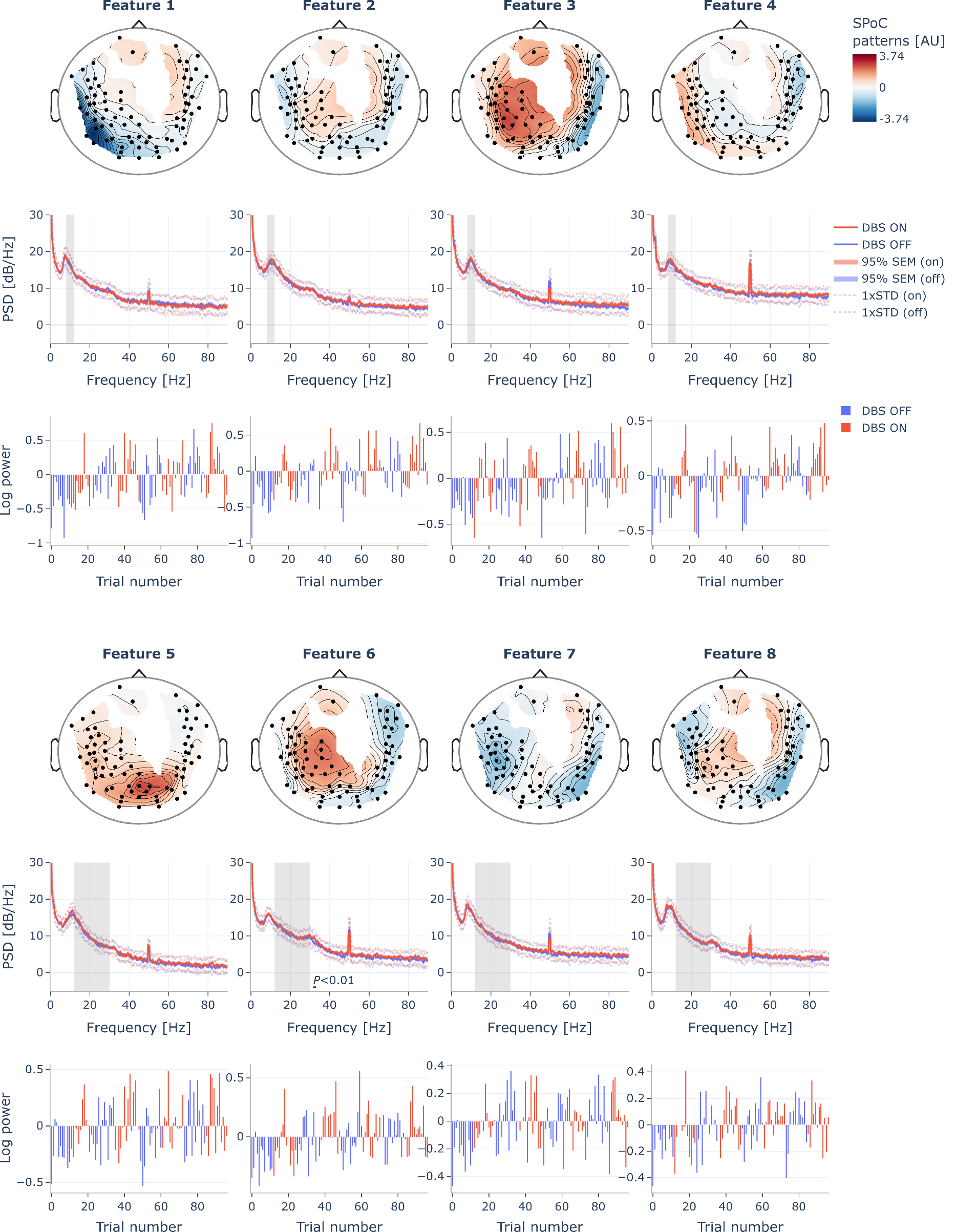}      
    \caption{EEG marker for $\text{S12}_2$. This figure showcases the eight selected filter bank source power comodulation features which are found by fitting our neural decoding pipeline for prediction of the CopyDraw scores. For each feature, the topological pattern, power spectral density (PSD), and the total band power of the spatially filtered signal per trial are shown. PSDs reflect the full spectrum of all trials grouped by DBS condition. A gray background highlights the frequency band the SPoC components (pattern) belong to. The log power per trial represents the frequency filtered (see gray background in PSD plot) and spatially filtered (SPoC filter) signal as is used as input for the final linear regression. Individual trials are colored according to their DBS condition. }
    \label{fig:apx_eeg_marker_S12_2}
\end{figure}

\begin{figure}
    \centering
    \includegraphics[width=\linewidth]{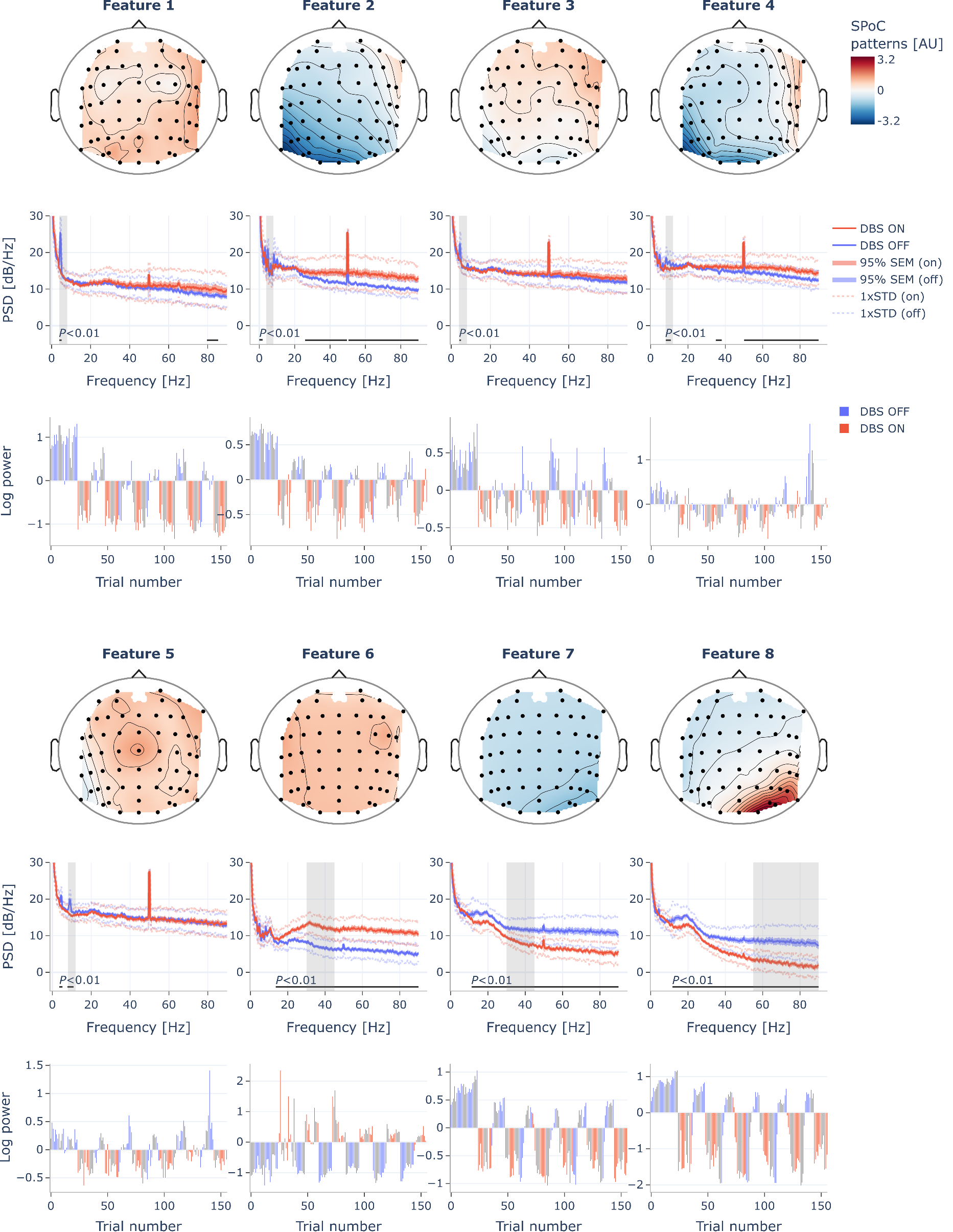}      
    \caption{EEG marker for $\text{S10}_c$. For details see caption of Figure~\ref{fig:apx_eeg_marker_S12_2}.}
    \label{fig:apx_eeg_marker_S10_c}
\end{figure}

\begin{figure}
    \centering
    \includegraphics[width=\linewidth]{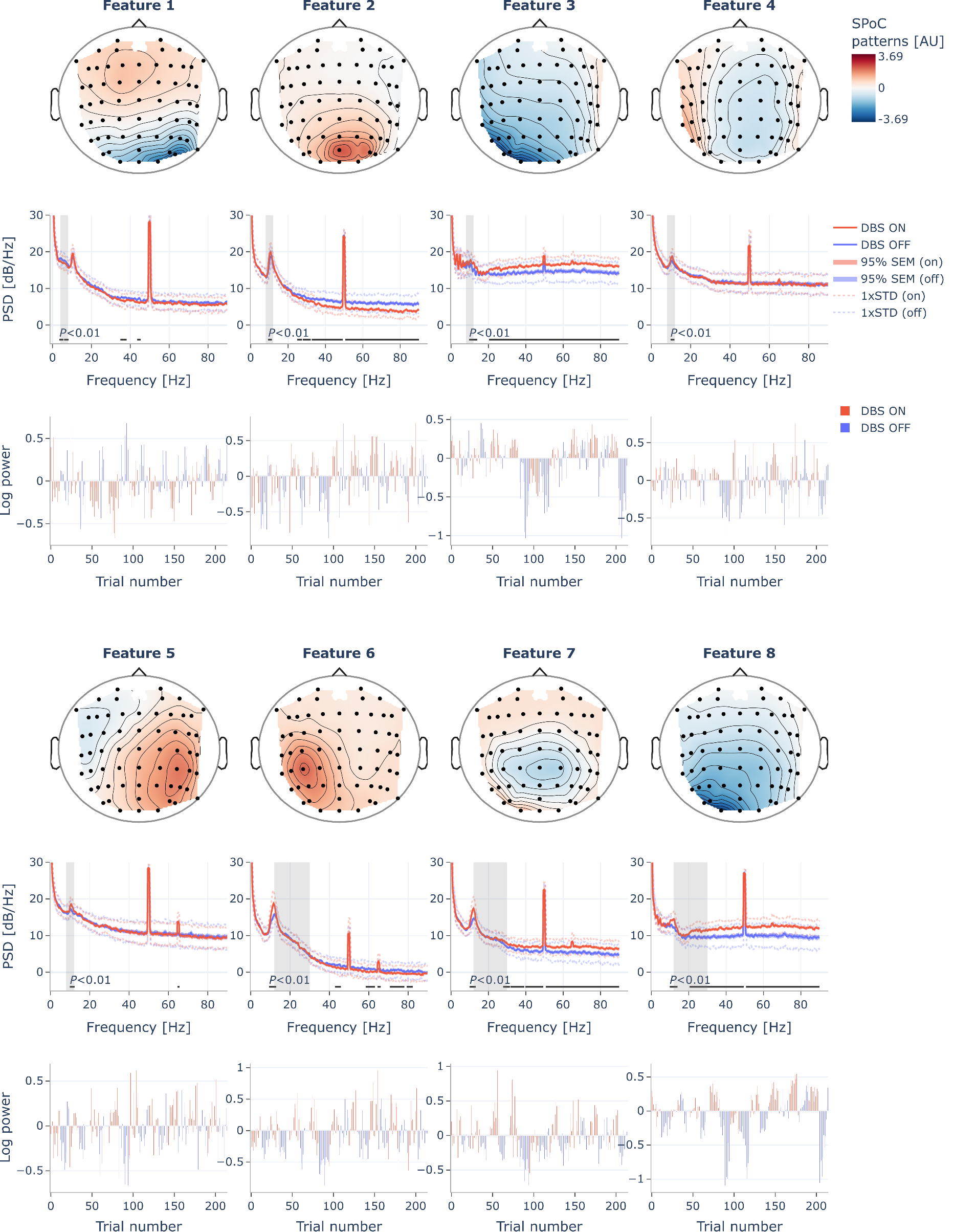}      
    \caption{EEG marker for $\text{S1}_c$. For details see caption of Figure~\ref{fig:apx_eeg_marker_S12_2}.}
    \label{fig:apx_eeg_marker_S1_c}
\end{figure}

\begin{figure}
    \centering
    \includegraphics[width=\linewidth]{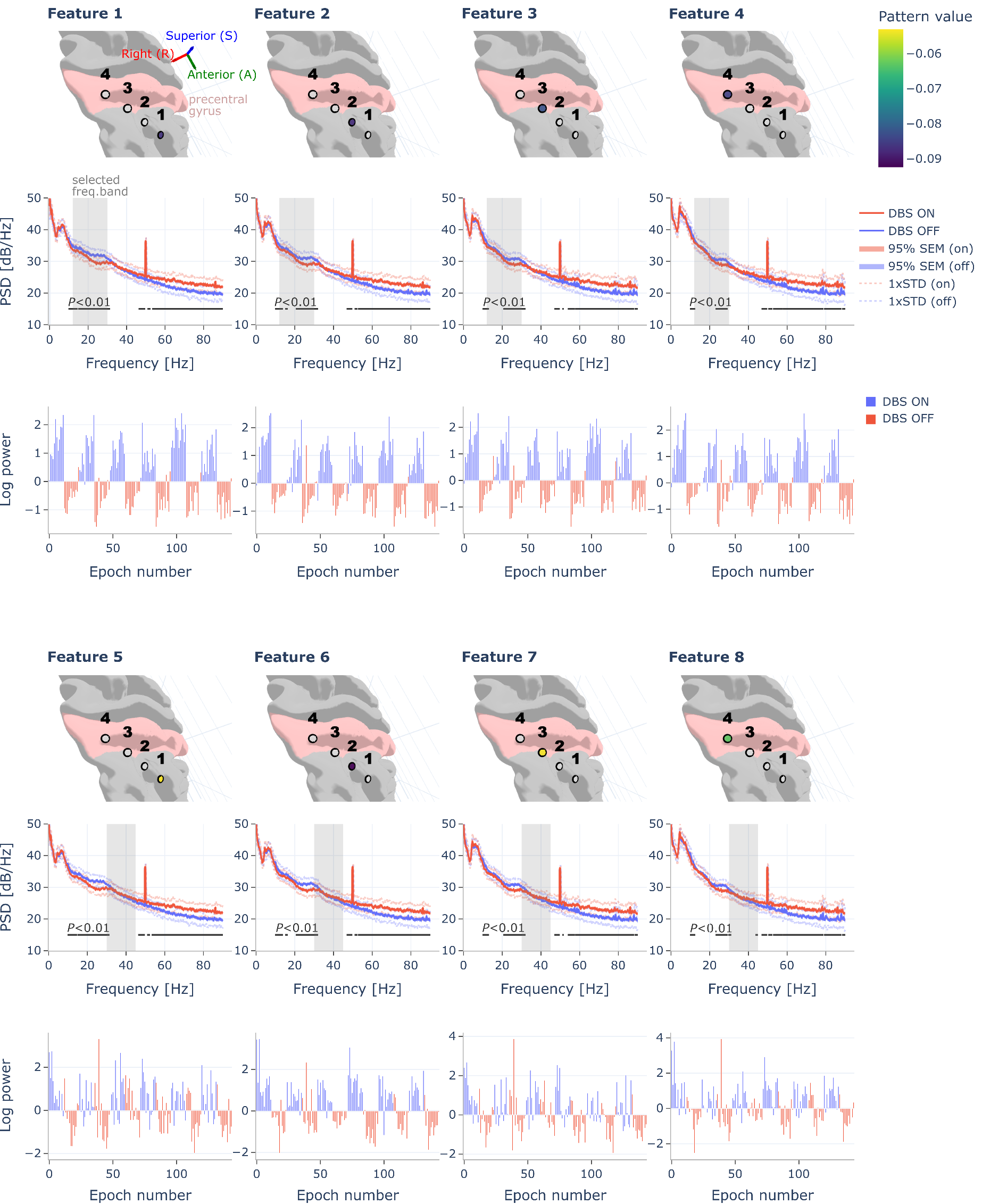}      
    \caption{ECoG marker for $\text{S17}_3$. Details are as described in the caption of Figure~\ref{fig:apx_eeg_marker_S12_2}, but with single ECoG channels and their weight pattern (linear model weights scaled by inverse covariance) shown instead of the SPoC patterns. The channels are shown over the fsaverage brain left hemisphere with the precentral gyrus highlighted in pink.}
    \label{fig:apx_ecog_marker}
\end{figure}

\begin{figure}
    \centering
    \includegraphics[width=0.7\linewidth]{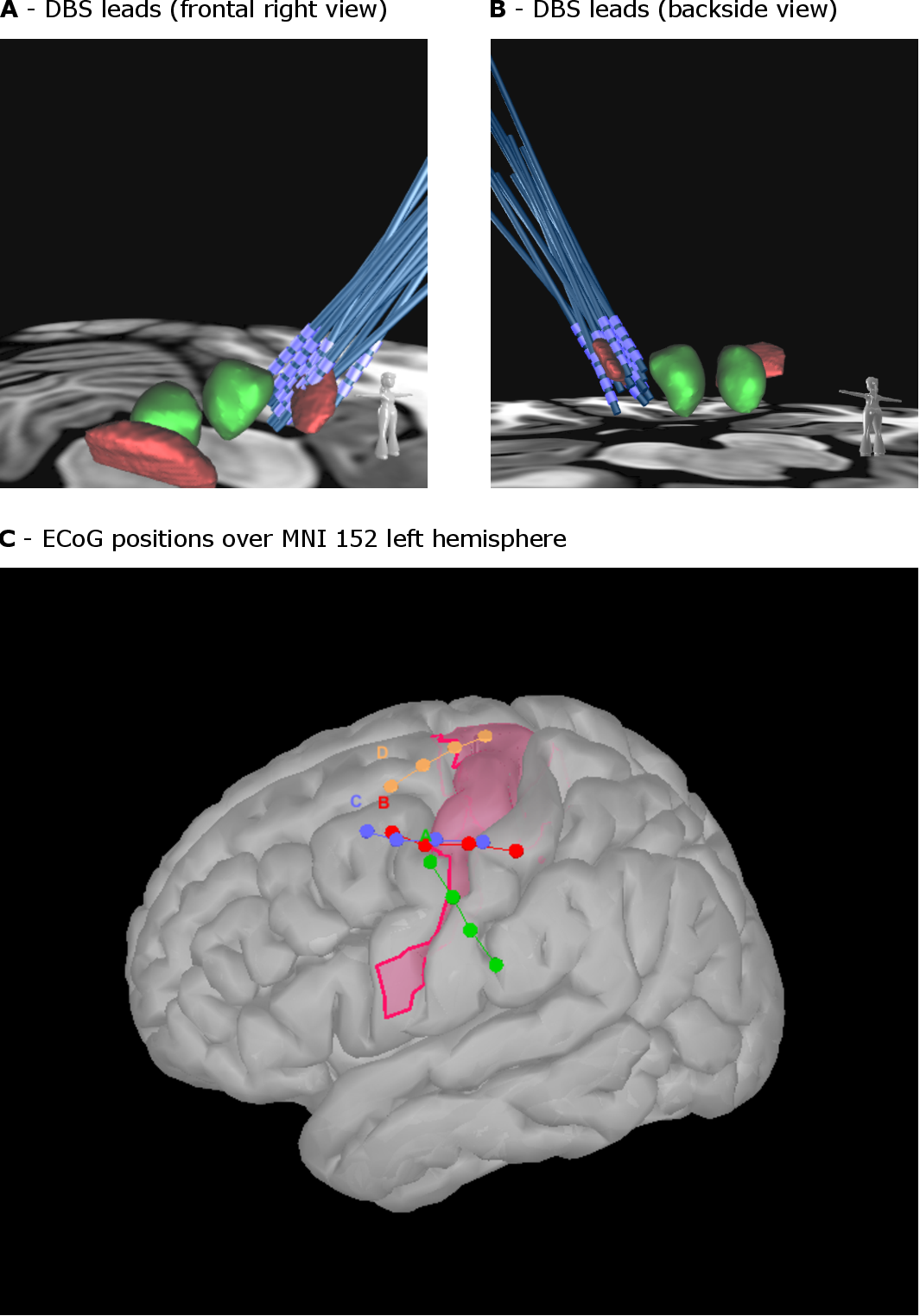}      
    \caption{DBS lead and ECoG positions. \textbf{A} and \textbf{B}: the frontal right and backside view of the left hemisphere DBS leads of the Freiburg cohort with acute session in common MNI 152 coordinates. Subthalamic nucleus (STN) and substantia nigra are shown in red and green respectively. A humanoid figure in the lower right serves as a reference for anatomical orientation. \textbf{C}: Positions of the ECoG strips over the left hemisphere of the MNI 152 brain. Green ECoG positions (A) refer to S14, red ECoG positions (B) refer to S15, blue ECoG positions (C) refer to S16, and orange ECoG positions (D) refer to S17. The motor area of the precentral gyrus is highlighted in pink.}
    \label{fig:apx_dbs_leads_and_ecog}
\end{figure}

\end{document}